\documentclass[preprint]{elsarticle}
\usepackage[utf8x]{inputenc}
\usepackage[T2A]{fontenc} 
\usepackage[russian,english]{babel}
\usepackage{amsmath,amssymb,amsfonts}
\usepackage{color}
\RequirePackage{titlesec}
\RequirePackage[titles]{tocloft}
\RequirePackage{graphicx}
	\graphicspath{{pictures/}}
\RequirePackage{epstopdf}
\RequirePackage{hyperref}

\begin{document}

\title{Numerical scheme for treatment of Uehling-Uhlenbeck equation for two-particle interactions in relativistic plasma}

\author[M,U,QO]{M. A. Prakapenia\corref{cor1}}
\ead{nikprokopenya@gmail.com}

\author[M,LTF]{I. A. Siutsou\corref{cor1}}
\ead{siutsou@icranet.org}

\author[M,P]{G. V. Vereshchagin}
\ead{veresh@icra.it}

\cortext[cor1]{Corresponding author}

\address[M]{ICRANet-Minsk, Institute of physics, National academy of sciences of Belarus\\ 220072 Nezale\v znasci Av. 68-2, Minsk, Belarus}
\address[U]{Department of Theoretical Physics and Astrophysics, Belarusian State University\\ 220030 Nezale\v znasci Av. 4, Minsk, Belarus}
\address[P]{ICRANet, 65122 Piazza della Repubblica, 10, Pescara, Italia}

\begin{abstract}

We present a new efficient method to compute Uehling-Uhlenbeck collision integral for all two-particle interactions in relativistic plasma with drastic improvement in computation time with respect to existing methods. Plasma is assumed isotropic in momentum space. The set of reactions consists of: Moeller and Bhabha scattering, Compton scattering, two-photon pair annihilation, and two-photon pair production, which are described by QED matrix elements. In our method exact energy and particle number conservation laws are fulfilled. Reaction rates are compared, where possible, with the corresponding analytical expressions and convergence of numerical rates is demonstrated.

\end{abstract}


\maketitle

\newpage
\section{Introduction}\label{ch2}

Relativistic plasma, for which $kT\geq mc^2$, where $k$ is Boltzmann constant, $c$ is speed of light, $m$ is electron mass, $T$ is temperature, is relevant in different branches of astrophysics. In the early universe ultrarelativistic electron-positron pairs contribute to the matter contents of the Universe \cite{weinberg2008cosmology}. X-ray and gamma-ray radiation from numerous astrophysical sources such as gamma-ray bursts \cite{1999PhR...314..575P,2010PhR...487....1R,2015PhR...561....1K}, active galactic nuclei \cite{2012AAT...27..557A, blandford2013active}, and X-ray binaries \cite{2006ARA&A..44..323F} points out to existence of relativistic electron-positron plasma in these objects.
The upcoming high-energy laser facilities aiming at generation of femtosecond laser pulses with intensity more than $10^{21}W/cm^2$ aim generation of relativistic plasma by interacting laser pulses. At present relativistic electron-positron jets are generated by interaction of laser pulses with condensed matter \cite{2015NatCo...6E6747S,0741-3335-53-1-015009,1742-6596-454-1-012016,PhysRevLett.102.105001}. 

%Computer simulations in plasma physics were started for the case of non-relativistic elecron-ion plasma, which can be treated as colissionless for a wide range of parameters. Most commonly used method for modeling of such a plasma is a particle-in-cell method (PIC) \cite{birdsall2004plasma,1983RvMP...55..403D,blum1989numerical,jardin2010computational}.
%Considering plasma as a set of individual particles interacting via self-consistently generated fields one approximates a subset of a real particles by one numerical 'macroparticle', therefore PIC method shrinks the computational time in a comparison with an exact N-body simulation.
%Modeling a Coulomb interactions in plasma one usually uses Monte-Carlo (MC) based methods \cite{SHERLOCK20082286,HUTHMACHER2016535,TURRELL2015144,BOBYLEV2013123},
% but there are also special PIC methods for such a process \cite{JONES1996169,LARSON2003123,LEMONS20091391}. 
%As a rule MC techniques are based on the random pairing of particles in close vicinity and the calculation of a scattering angle due to the interaction. Small-angle Coulomb collisions which allow small energy and momentum transfer are often described in diffusion approximation by the Fokker-Planck equation \cite{1981phki.book.....L}. The principal feature of relativistic plasma is a presence of pair creation and pair annihilation processes, which are often included in MC based models \cite{0741-3335-53-1-015009,1742-6596-454-1-012016}. 

Solving the Boltzmann equations with collision integral containing a quantum cross-section represents the most general and complete method to describe a behavior of relativistic plasma \cite{vereshchagin2017relativistic,cercignani2012relativistic,groot1980relativistic}. 
The one-particle distribution function (DF) is defined on a seven dimensional space, three dimensions for the physical space and three dimensions for the momentum space, and one dimension for the time. Thus one has a multidimensional problem which is a real challenge from the computational point of view. Beside the dimensionality problem, there are other difficulties which are related to kinetic equations in general \cite{1990JMP....31..245B,1991RvMaP...3..137B}. 
Our main goal in this paper is to tackle the challenge associated with the calculation of the collision integral, dealing with two key issues. First, the computational cost related to the evaluation of the collision operator involving multidimensional integrals which should be solved in each point of the coordinate space. Second, the presence of multiple scales requires the development of adapted numerical schemes capable of solving stiff dynamics. Different deterministic approaches are used to tackle collision integral from a numerical point of view: finite volume, semi-Lagrangian and spectral schemes \cite{1991RvMaP...3..137B,dimarco_pareschi_2014,2006MaCom..75.1833M,DIMARCO2017,WU201327}. While the deterministic methods could normally reach high order of accuracy, the probabilistic ones, such as Monte-Carlo (MC) method, are often faster.

MC methods are traditionally used to model Coulomb interactions in non-relativistic plasma\cite{SHERLOCK20082286,HUTHMACHER2016535,TURRELL2015144,BOBYLEV2013123}. As a rule MC techniques are based on the random pairing of particles in close vicinity and the calculation of a scattering angle due to the interaction. Small-angle Coulomb collisions which allow small energy and momentum transfer are often described in diffusion approximation by the Fokker-Planck equation \cite{1981phki.book.....L}. The principal feature of relativistic plasma is a presence of pair creation
and pair annihilation processes, which are often included in MC based models \cite{0741-3335-53-1-015009,1742-6596-454-1-012016}. However Fokker-Planck approximation is no longer valid in relativistic plasma \cite{2009PhRvD..79d3008A}.

Classical Boltzmann equation does not take into account quantum statistics of particles. The generalization of classical Boltzmann equation including quantum corrections is Uehling-Uhlenbeck (U-U) equation, which contains additional Pauli blocking and Bose enhancement multipliers that give rise to equilibrium solution with Bose-Einstein and Fermi-Dirac distributions \cite{1934PhRv...46..917U,1933PhRv...43..552U}. The main problem of the MC methods in application to U-U equations is that total reaction rate is unknown as distribution function is unknown too. Compensation methods include smoothing of the delta-function distribution of MC-particles over cells in the phase space, but it suffers from a large number of simulation particles and cells needed to reproduce Bose-Einstein steady state distribution. Spectral methods based on the Fourier transformation of the velocity distribution function require very dense computational grid to reach high accuracy \cite{2017JCoPh.330.1010Y,Hu2015,huying12,PhysRevE.68.056703,2010arXiv1009.3352F}.
Process-oriented approach to the U-U collision integral presented in this work allows one to get high accuracy results with low computational cost.

In this paper we further develop the method first used in the work \cite{2004ApJ...609..363A}. This method was successfully applied to follow the thermalization of relativistic plasma \cite{2007PhRvL..99l5003A,2009AIPC.1111..344A,2009PhRvD..79d3008A,2010AIPC.1205...11A} and to investigate thermalization timescales for an electron-positron plasma \cite{2010PhRvE..81d6401A}. In section \ref{ch3} we recall Boltzmann and UU equations and present usual scheme of their analytic treatment. Section \ref{ch4} is devoted to the description of our numerical scheme, while section \ref{ch5} shows comparison between our code results and known analytic formulae for non-degenerate case. Conclusion follows.

\newpage
\section{Formulation}\label{ch3}

The Boltzmann equation governs an evolution of one-particle distribution function $f(\mathbf x,\mathbf p,t)$. We assume that plasma is homogeneous and isotropic in coordinate space and isotropic in momentum space, thus distribution function depends on absolute value of momentum (energy) and time. DF is normalized on particles concentration, so that $n=\int f(\mathbf p,t)d^3p$.\\
Consider an interaction of two initial particles of type I and II which are in states 1 и 2, correspondingly, and creation of two final particles of type III и IV which are in states 3 и 4, correspondingly. Let us image the process by the following scheme:
\begin{equation}\label{2pdir}
I_1 + II_2 \rightarrow III_3 + IV_4.
\end{equation}
The corresponding inverse process is:
\begin{equation}\label{2pinv}
III_3 + IV_4\rightarrow I_1 + II_2.
\end{equation}
If every particle has momentum $p_i$, which lies in interval $d^3p_i$, then a number of interactions in unit time and unit space volume is:
\begin{equation}
w(p_1,p_2;p_3,p_4)f_{I} f_{II} d^3p_1d^3p_2d^3p_3d^3p_4,
\end{equation}
function $w$ is called a transition rate for a given reaction. \\
An effective cross-section is defined by the formula:
\begin{equation}
d\sigma=\frac{w}{v}d^3p_3 d^3p_4,
\end{equation}
where $v=c\epsilon_1^{-1}\epsilon_2^{-1}\sqrt[]{\left( \epsilon_1\epsilon_2-(\mathbf{p_1}\mathbf{p_2})c^2 \right)^2-(m_1 m_2 c^4)^2}$ is a relative velocity of particles. \\
In quantum field theory an expression for interaction cross-section is:
\begin{equation}\label{dsigma}
d\sigma=\frac{\hbar^2 c^6}{(2\pi)^2}\frac{1}{v}\frac{|M_{if}|^2}{16\epsilon_1\epsilon_2\epsilon_3\epsilon_4}
\delta(\epsilon_1+\epsilon_2-\epsilon_3-\epsilon_4)\delta(\mathbf p_1+\mathbf p_2-\mathbf p_3-\mathbf p_4)d^3p_3d^3p_4,
\end{equation}
where $|M_{if}|$ are a matrix elements calculated with a methods of quantum field theory. \\
Comparing two last formulas one can derive the following expression for transition rate:
\begin{equation}\label{transw}
w(p_3,p_4;p_1,p_2)=\frac{\hbar^2 c^6}{(2\pi)^2}\frac{|M_{if}|^2}{16\epsilon_1\epsilon_2\epsilon_3\epsilon_4}\delta(\epsilon_1+\epsilon_2-\epsilon_3-\epsilon_4)
\delta(\mathbf p_1+\mathbf p_2-\mathbf p_3-\mathbf p_4),
\end{equation}
Now let us write the Boltzmann equation for DF of particle I for a given process:
\begin{multline}\label{StfI}
\dot{f_I}=\int d^3p_2 d^3p_3 d^3p_4 [w(p_3,p_4;p_1,p_2)f_{III}(\mathbf p_3,t)f_{IV}(\mathbf p_4,t)\\
-w(p_1,p_2;p_3,p_4)f_{I}(\mathbf p_1,t)f_{II}(\mathbf p_2,t)],
\end{multline}
where a dot denotes time derivative. Equations for particle DFs of remaining types can be derived by the corresponding replacement of indices.\\
Specifically, for a scattering with $I=III$ and $II=IV$ the inverse process is the same as the direct one since pairs of indices $(1,2)$ and $(3,4)$ can be interchanged. The relation $w(p_3,p_4;p_1,p_2)=w(p_1,p_2;p_3,p_4)$ holds for all processes listed in Table \ref{2pptable}.
The right hand side of the Boltzmann equation is a collision integral denoted as St$f_I$. The first term in collision integral describes particle outcome and the second term describes particle income, we will denote it as $\text{St}^-f$ and $\text{St}^+f$, respectively.\\

\begin{table}
\caption{Two-particle processes in electron-positron-photon plasma\label{2pptable}}
\center
\begin{tabular}[c]{|l|l|l|l|l|}
\hline
Process (q) & I & II & III & IV \\
\hline
Compton Scattering (CS) & $e^{\pm}$ & $\gamma$ & $e^{\pm}$ & $\gamma$\\
\hline
Bhabha Scattering (BS) & $e^{\pm}$ & $e^{\mp}$ & $e^{\pm}$ & $e^{\mp}$\\
\hline
M{\o}ller Scattering (MS) & $e^{\pm}$ & $e^{\pm}$ & $e^{\pm}$ & $e^{\pm}$\\
\hline
Pair Annihilation (PA) & $e^{-}$ & $e^{+}$ & $\gamma$ & $\gamma$\\
\hline
Pair Creation (PC) & $\gamma$ & $\gamma$ & $e^{-}$ & $e^{+}$\\
\hline
\end{tabular}
\end{table}

The generalization of Boltzmann equation for the case of particles obeying quantum statistics is U-U equation. For the particle $I$ in the state $1$ U-U equation has the following form:
\begin{multline}  \label{uustfI}
\dot{f_I} =\int d^{3} \mathbf{p}_{2} d^{3} \mathbf{p}_{3} d^{3} \mathbf{p}_{4}  \\
\shoveleft{\ \times\biggl[w(p_3,p_4;p_1,p_2)
f_{III}(\mathbf{p}_{3},t) f_{IV}(\mathbf{p}_{4},t)
\left( 1+\eta \frac{f_{I}(\mathbf{p}_{1},t)}{2 h^{-3}}\right)\left( 1+\eta \frac{f_{II}(\mathbf{p}_{2},t)}{2 h^{-3}}\right)}\\
\shoveleft{\quad -w(p_1,p_2;p_3,p_4)
f_{I}(\mathbf{p}_{1},t)f_{II}(\mathbf{p}_{2},t)}
\left(1+\eta \frac{f_{III}(\mathbf{p}_{3},t)}{2 h^{-3}}\right) \left(1+\eta \frac{f_{IV}(\mathbf{p}_{4},t)}{2 h^{-3}}\right) \biggr],
\end{multline}
where $\eta$ is defined through
\begin{equation}
    \eta=
    \begin{cases}
        +1, & \text{for Bose-Einstein statistics,}\\
 	  -1, & \text{for Fermi-Dirac statistics,}\\
         0, & \text{for Maxwell-Boltzmann statistics.}
    \end{cases}
\end{equation}
When incoming or outgoing particles coincide ($I=II$ and/or $III=IV$) quantum indistinguishability gives the term $\frac12$ in front of the corresponding
outcome and income terms, see e.g. \cite{1973rela.conf....1E}, \cite{groot1980relativistic}.

For numerical evaluation phase space is divided into zones, in calculations we approximate continuous DF by its averaging over each zone (see Eq.~\eqref{Ydef}). For this purpose we add an integral over $\mathbf{p}_{1}$ in UU equation \ref{uustfI}, the RHS of resulting equation will have the same form for each particle type differing only by sign and its integration limits:
\begin{multline}\label{uustfIint}
\pm \int d^{3} \mathbf{p}_{1} d^{3} \mathbf{p}_{2} d^{3} \mathbf{p}_{3} d^{3} \mathbf{p}_{4}  \\
\shoveleft{\ \times\biggl[w(p_3,p_4;p_1,p_2)
f_{III}(\mathbf{p}_{3},t) f_{IV}(\mathbf{p}_{4},t)
\left( 1+\eta \frac{f_{I}(\mathbf{p}_{1},t)}{2 h^{-3}}\right)\left( 1+\eta \frac{f_{II}(\mathbf{p}_{2},t)}{2 h^{-3}}\right)}\\
\shoveleft{\quad -w(p_1,p_2;p_3,p_4)
f_{I}(\mathbf{p}_{1},t)f_{II}(\mathbf{p}_{2},t)}
\left(1+\eta \frac{f_{III}(\mathbf{p}_{3},t)}{2 h^{-3}}\right) \left(1+\eta \frac{f_{IV}(\mathbf{p}_{4},t)}{2 h^{-3}}\right) \biggr],
\end{multline}
where the upper sign corresponds to particle type I and II and the lower sign corresponds to particle type III and IV.

Evaluating collision integral in the framework of reaction-oriented approach we use one expression \eqref{uustfIint} and distribute the result to each particle type according to integration limits in \eqref{uustfIint}.

In this paper we deal with all two-particle QED processes in relativistic plasma, which are collected in Table \ref{2pptable}. The exact QED matrix elements for these processes can be found in the standard textbooks, e.g. \cite{2003spr..book.....G,1982els..book.....B}.\\

%\begin{table}\label{2pcoeftable}
%\caption{Emission and absorbsion coefficients for two-particle processes in electron-positron-photon plasma}
%\center
%\begin{tabular}[c]{|l|l|l|l|l|l|l|}
%\hline
%q & $\eta_{-}$ & $\chi_{-}$ & $\eta_{+}$ & $\chi_{+}$ & $\eta_{\gamma}$ & $\chi_{\gamma}$ \\
%\hline
%CS & $\checkmark$ & $\checkmark$ & $\checkmark$ & $\checkmark$ & $\checkmark$ & $\checkmark$ \\
%\hline
%BS & $\checkmark$ & $\checkmark$ & $\checkmark$ & $\checkmark$ & $\times$ & $\times$ \\
%\hline
%MS & $\checkmark$ & $\checkmark$ & $\checkmark$ & $\checkmark$ & $\times$ & $\times$ \\
%\hline
%PA & $\times$ & $\checkmark$ & $\times$ & $\checkmark$ & $\checkmark$ & $\times$ \\
%\hline
%PC & $\checkmark$ & $\times$ & $\checkmark$ & $\times$ & $\times$ & $\checkmark$ \\
%\hline
%\end{tabular}
%\end{table} \\
%\newpage
Let us make a notice connected with a conservation laws for interacting particles. Energy and momentum conservations read
\begin{gather}\label{2pconslaws}
\hat\varepsilon = \varepsilon_{1} + \varepsilon_{2}=\varepsilon_{3} +
\varepsilon_{4},\qquad \hat{\mathbf p}=\mathbf p_{1} + \mathbf p_{2}=\mathbf p_{3} + \mathbf
p_{4}.
\end{gather}
There are 4 delta-functions in Eq.~\eqref{transw} representing conservation of energy and momentum \eqref{2pconslaws}. Three integrations over momentum of particle $III$ can be performed immediately
\begin{equation}
    \int d\mathbf p_3 \delta^3(\mathbf{p}_{1}+\mathbf{p}_{2}-\mathbf{p}_{3}-\mathbf{p}_{4})\longrightarrow 1.
\end{equation}
In the integration over energy $\varepsilon_{4}$ of particle $IV$ it is necessary to take into account that $\varepsilon_3$ is now a function of energy and angles of particles $I$ and $II$, as well as angles of particle $IV$, so we have
\begin{equation}
    \int d\varepsilon_4 \delta(\varepsilon_{1}+\varepsilon_{2} -\varepsilon_{3}-\varepsilon_{4}) \longrightarrow
        \frac{1}{1-(\beta_{3}/\beta_{4})\mathbf n_{3}\cdot\mathbf n_{4}},
\end{equation}
where $\mathbf n=\mathbf p/p$ is the unit vector in the direction of particle momentum, $p=|\mathbf p|=\sqrt{(\varepsilon/c)^2 -m^2c^2}$ is the absolute value of particle momentum, $\beta=pc/\varepsilon$, and a dot denotes scalar product of 3-vectors.\\
We use spherical coordinates in momentum space: $\{\varepsilon, \mu ,\phi \}$, $\mu=\cos\vartheta$, where $\varepsilon$ is the particle energy, and $\vartheta$ and $\phi$ are polar and azimuthal angles, respectively. Then energy and angles of particle $III$ and energy of particle $IV$ follow from energy and momentum conservation \eqref{2pconslaws} and relativistic energy-momentum relation, namely
\begin{gather}\label{KinematicsStart}
    \varepsilon_4=c\sqrt{p_4^2+m_{IV}^2c^2},\qquad \mathbf p_4=p_4 \mathbf n_4,\\
    \varepsilon_3=\hat\varepsilon-\varepsilon_4,\qquad
    \mathbf p_3=\hat{\mathbf p}-\mathbf p_4,\nonumber\\
    \mathbf n_3=\frac{\mathbf p_3}{p_3},\qquad \mathbf n_4=\frac{\mathbf p_4}{p_4},\\
    \mathbf{n}_3=\left(\sqrt{1-{\mu_3}^2}\cos\phi_3, \sqrt{1-{\mu_3}^2}\sin\phi_3,
    \mu_3\right),\\
    \mathbf{n}_4=\left(\sqrt{1-\mu_4^2}\cos\phi_4, \sqrt{1-\mu_4^2}\sin\phi_4,
    \mu_4\right),\\
    p_4=\frac{AB\pm\sqrt{A^2+4m_{IV}^2c^2(B^2-1)}}{2(B^2-1)},\label{Kinematics}\\
    A=\frac{c}{\hat\varepsilon}[\hat{p}^2+(m_{III}^2-m_{IV}^2)c^2]
    -\frac{\hat\varepsilon}{c},\qquad
    B=\frac{c}{\hat\varepsilon}\mathbf n_4\cdot\hat{\mathbf p}.\nonumber
\\
    \mathbf n_{3}\cdot\mathbf n_{4}=\mu_3\mu_4+\sqrt{(1-\mu_3^2)(1-\mu_4^2)}\cos(\phi_3-\phi_4).
    \label{KinematicsEnd}
\end{gather}

Then we introduce these relations into collision integral \eqref{uustfIint}. We also use spherical symmetry in momentum space to fix angles of the particle $I$: $\mu_1=1,\phi_1=0$, and to perform the integration over azimuthal angle of particle $II$: $\int d\phi_2\longrightarrow2\pi$, setting $\phi_2=0$ in the remaining integrals. Then final expression for collision integral is
\begin{multline}\label{stfI1}
\text{St}f_I=\frac{\hbar^2}{32\pi}\int d\varepsilon_2 d\mu_2 \ d\mu_4 d\phi_4
\frac{p_2 p_4|M_{fi}|^2}{\varepsilon_{1}\varepsilon_{3}[1-(\beta_{3}/\beta_{4})\mathbf n_{3}\cdot\mathbf n_{4}]}\\
\shoveleft{\ \times\biggl[f_{III}(\varepsilon_{3},t) f_{IV}(\varepsilon_{4},t) \left( 1+\eta \frac{f_{I}(\varepsilon_{1},t)}{2 h^{-3}}\right)
\left( 1+\eta \frac{f_{II}(\varepsilon_{2},t)}{2 h^{-3}}\right)}\\
\shoveleft{\quad -f_{I}(\varepsilon_{1},t)f_{II}(\varepsilon_{2},t)}\left(1+\eta\frac{f_{III}(\varepsilon_{3},t)}{2 h^{-3}}\right)
\left(1+\eta\frac{f_{IV}(\varepsilon_{4},t)}{2 h^{-3}}\right) \biggr].
\end{multline}

For numerical integration, however, another expression is proved useful
\begin{multline}\label{stfI2}
\int d\varepsilon_1 \text{St}f_I =\frac{\hbar^2}{32\pi}\Biggl[ \int d\varepsilon_3\ d\varepsilon_4 d\mu_4 \ d\mu_2 d\phi_2
\frac{p_2 p_4|M_{fi}|^2}{\varepsilon_{1}\varepsilon_{3}\,[1-(\beta_{1}/\beta_{2})\mathbf n_{1}\cdot\mathbf n_{2}]}\\
    \times f_{III}(\varepsilon_{3},t) f_{IV}(\varepsilon_{4},t)\left(1+\eta\frac{f_{I}(\varepsilon_{1},t)}{2 h^{-3}}\right) \left(1+\eta \frac{f_{II}(\varepsilon_{2},t)}{2 h^{-3}}\right)\\
-\int d\varepsilon_1 \ d\varepsilon_2 d\mu_2 \ d\mu_4 d\phi_4
\frac{p_2 p_4|M_{fi}|^2}{\varepsilon_1 \varepsilon_{3}\,[1-(\beta_{3}/\beta_{4})\mathbf n_{3}\cdot\mathbf n_{4}]}\\
\shoveleft{\ \times f_{I}(\varepsilon_1,t)f_{II}(\varepsilon_{2},t)}\left(1+\eta\frac{f_{III}(\varepsilon_{3},t)}{2 h^{-3}}\right) \left(1+\eta \frac{f_{IV}(\varepsilon_{4},t)}{2 h^{-3}}\right)\Biggr],
\end{multline}
where the first term is expressed in the form ready for replacement by the sum over incoming particles $III$ and $IV$. In this term $\varepsilon_1,\mu_1,\phi_1,\varepsilon_2$ are given by relations \eqref{Kinematics} with indices exchange $1\leftrightarrow3$, $2\leftrightarrow4$, $I\leftrightarrow III$, $II\leftrightarrow IV$.

This collision integral of any of two-particle processes is a four-dimensional integral in momentum space. In Sec.~\ref{ch4} we show how such integral is computed numerically on finite grid.

Here we note that in the case of homogeneous and isotropic pair plasma one has to satisfy only two conservation laws, namely of energy and particle number. Momentum conservation should be added for nonisotropic in momentum space DF, see e.g.~\cite{BENEDETTI2013206}. In our method electric charge is conserved due to conservation of particles because we use between cell interpolation for the same kind of particles described in the next Section. 
\newpage
\section{Numerical Scheme}\label{ch4}

%The main difficulty arising in computation of collision integrals in comparison
%with previous works \cite{2007PhRvL..99l5003A, 2009PhRvD..79d3008A,
%Aksenov2009} is that particle emission and absorbtion coefficients contain not
%only distribution functions of incoming particles, but also those of outgoing
%particles. Therefore we adopt a different approach which we refer to as
%"\emph{reaction-oriented}" instead of "particle-oriented" one used earlier.

The phase space is divided in zones. The zone $\Omega^\alpha_{a,j,k}$ for particle specie $\alpha$ corresponds to energy $\varepsilon_a$, cosine of polar angle $\mu_j$ and azimuthal angle $\phi_k$, where indices run in the following ranges $1\leq a\leq a_{\mathrm{max}}$, $1\leq j\leq j_{\mathrm{max}}$, and $1\leq k\leq k_{\mathrm{max}}$. The zone boundaries are $\varepsilon_{a\mp1/2}$, $\mu_{j\mp1/2}$, $\phi_{k\mp1/2}$. The length of the $a$-th energy zone $\Omega^\alpha_{a}$ is $\Delta\varepsilon_{a} \equiv \varepsilon_{a+1/2} - \varepsilon_{a-1/2}$. On finite grid $f_\alpha$ does not depend on $\mu$ and $\phi$, and number density of particle $\alpha$ in zone $a$ is
\begin{multline}\label{Ydef}
Y^\alpha_{a}(t)=4\pi\int_{\varepsilon_{a-1/2}}^{\varepsilon_{a+1/2}}c^{-3}\varepsilon\sqrt{\varepsilon^2-m_\alpha^2c^4}
f_\alpha(\varepsilon,t)d\varepsilon\\= 4\pi c^{-3}\varepsilon_a\sqrt{\varepsilon_a^2-m_\alpha^2c^4} f_\alpha(\varepsilon_a,t)\Delta\varepsilon_a.
\end{multline}
In this variables discretized U-U equation for particle $I$ and energy zone $a$ reads
\begin{equation}\label{discrBoltzmann}
    \frac{d Y^\alpha_a(t)}{dt}=\sum \left[\text{St}^+Y^{I}_a -\text{St}^-Y^{I}_a \right],
\end{equation}
where the sum is taken over all processes involving particle $I$. Coefficients of particles income and outcome on the grid are obtained by integration of \eqref{stfI2} for two-particle processes over the zone. The corresponding integrals are replaced by sums on the grid.
For instance, coefficient of particle $I$ outcome in two-particle process \eqref{2pdir} is
\begin{multline}\label{YrateI}
    \text{St}^-Y^{I}_a=\frac{\hbar^2c^{4}}{8(4\pi)^2} \sum_{b,j,s,k}\Delta\mu^{II}_{j}\Delta\mu^{IV}_{s} \Delta\phi^{IV}_{k} |M_{fi}|^2 \frac{p_4}{\varepsilon_{3}[1-(\beta_{3}/\beta_{4})\mathbf n_{3}\cdot\mathbf n_{4}]}\times \\
\times  \frac{Y^{I}_{a}(t)}{\varepsilon^{I}_{a}}\frac{Y^{II}_{b}(t)}{\varepsilon^{II}_{b}}
    \times\left[1+\eta \frac{Y^{III}_c(t)}{\bar Y^{III}_c}\right]\left[1+\eta \frac{Y^{IV}_d(t)}{\bar Y^{IV}_d}\right],
\end{multline}
and coefficient of particle $I$ income in process (\ref{2pinv}) from integration of \eqref{stfI2} is
\begin{multline}\label{YrateII}
    \text{St}^+Y^{I}_a=\frac{\hbar^2c^{4}}{8(4\pi)^2} \sum_{c,d,j,s,k} C_a(\varepsilon_1) \Delta\mu^{IV}_{j}\Delta\mu^{II}_{s} \Delta\phi^{II}_{k} |M_{fi}|^2 \frac{p_2}{\varepsilon_{1}[1-(\beta_{1}/\beta_{2})\mathbf n_{1}\cdot\mathbf n_{2}]} \times \\
\times  \frac{Y^{III}_{c}(t)}{\varepsilon^{III}_{c}}\frac{Y^{IV}_{d}(t)}{\varepsilon^{IV}_{d}}
    \times\left[1+\eta \frac{Y^{I}_a(t)}{\bar Y^{I}_a}\right]\left[1+\eta \frac{Y^{II}_b(t)}{\bar Y^{II}_b}\right],
\end{multline}
where $\bar Y^\alpha_{a}=4\pi\int_{\varepsilon_{a-1/2}}^{\varepsilon_{a+1/2}}
        c^{-3}\varepsilon\sqrt{\varepsilon^2-m_\alpha^2c^4}\ (2 h^{-3})d\varepsilon
        = 8\pi(hc)^{-3}\varepsilon_a\sqrt{\varepsilon_a^2-m_\alpha^2c^4} \Delta\varepsilon_a,$
and
\begin{equation}\label{csol}
    C_a(\varepsilon_1)=
    \begin{cases}
        \dfrac{\varepsilon_{a}-\varepsilon_1}{
            \varepsilon_{a}-\varepsilon_{a-1}}, &
        \varepsilon_{a-1} < \varepsilon_1 < \varepsilon_{a},\\[2.5ex]
        \dfrac{\varepsilon_{a+1}-\varepsilon_1}{
            \varepsilon_{a+1}-\varepsilon_{a}}, &
        \varepsilon_{a} < \varepsilon_1 < \varepsilon_{a+1},\\[2.5ex]
        0, & \text{otherwise.}
    \end{cases}
\end{equation}
In integration of (\ref{stfI2}) over the zone one can integrate out the $\delta$-function $\int \delta(\varepsilon_1-\varepsilon)d\varepsilon_1
\longrightarrow 1$. However, when energies of incoming particles are fixed on the grid, the energies of outgoing particles are not on the grid. Hence an interpolation \eqref{csol} is adopted, which enforces the exact number of particles and energy conservation in each two-particle process due to redistribution of outgoing particle $\alpha$ with energy $\varepsilon$ over two energy zones $\Omega^{\alpha}_{n}, \Omega^{\alpha}_{n+1}$ with $\varepsilon_{n} < \varepsilon < \varepsilon_{n+1}$. Further we denote this technique as particle splitting.

The redistribution of final particles should also satisfy requirements of quantum statistics. Therefore if a process occurs, when fermionic final particle should be distributed over the quantum states which are fully occupied, such process should be forbidden. Thus we introduce the Bose enhancement/Pauli blocking coefficients in (\ref{YrateI}) and (\ref{YrateII}) as
\begin{gather}
    \left[1+\eta \frac{Y^{\alpha}_a(t)}{\bar Y^{\alpha}_a}\right]=
    \min\left(1+\eta \frac{Y^{\alpha}_{n}(t)}{\bar Y^{\alpha}_{n}},
    1+\eta \frac{Y^{\alpha}_{n+1}(t)}{\bar Y^{\alpha}_{n+1}}\right).
\end{gather}
The sum over angles $\mu_j, \mu_s, \phi_k$ can be found once and for all at the
beginning of the calculations. We then store in the program for each set of the
incoming and outgoing particles the corresponding terms and redistribution
coefficients given by Eq.~(\ref{csol}).

Representation of discretized collisional integral for particle $I$ and energy zone $a$ in processes \eqref{2pdir},
\eqref{2pinv} is
\begin{multline}\label{BoltzmannFinal}
      \dot{Y}^I_a=
    \sum P_{abcd}\times Y^{III}_{c}(t) Y^{IV}_{d}(t)
    \times \left[1+\eta \frac{Y^{I}_{a}(t)}{\bar Y^{I}_{a}}\right]
    \left[1+\eta \frac{Y^{II}_{b}(t)}{\bar Y^{II}_{b}}\right]\\
-\sum R_{abcd}\times Y^{I}_{a}(t) Y^{II}_{b}(t)
    \times \left[1+\eta \frac{Y^{III}_{c}(t)}{\bar Y^{III}_{c}}\right]
    \left[1+\eta \frac{Y^{IV}_{d}(t)}{\bar Y^{IV}_{d}}\right],
\end{multline}
where constant coefficients $P,R$ are obtained from the summation over angles in the sums \eqref{YrateI}, \eqref{YrateII}. In the nondegenerate case of Boltzmann equation the indices $b$ in the first sum and $c,d$ in the second sum become dummy, equation \eqref{BoltzmannFinal} can be partially summed and takes the following form:
\begin{gather}\label{BoltzmannFinal2}
      \dot{Y}^I_a=
    \sum P_{acd}\times Y^{III}_{c}(t) Y^{IV}_{d}(t)
    -\sum R_{ab}\times Y^{I}_{a}(t) Y^{II}_{b}(t),
\end{gather}
where $P_{acd}=\sum_{b} P_{abcd},\ R_{ab}=\sum_{c,d} R_{abcd}$. The last quantity is essentially \emph{reaction rate} usually used for description of binary processes and simply connected to the total cross section.

The full U-U equation \eqref{BoltzmannFinal} contains similar sums for all processes from Table \ref{2pptable}. Each individual term in these sums appears in the system of discretized equations four times in emission and absorption coefficients for each particle entering a given process. Then each term can be computed only once and added to all
corresponding sums, that is the essence of our \emph{"reaction-oriented" approach} \cite{ivanphd,2015AIPC.1693g0007S}.

We point out that unlike classical Boltzmann equation for binary interactions
such as scattering, more general interactions are typically described by four
collision integrals for each particle that appears both among incoming and
outgoing particles. 
\newpage
\section{Numerical results}\label{ch5}

The results of numerical calculations are presented below. As all known analytical expressions for reaction rates in relativistic plasma concern nondegenerate case, here we compare our results for collision integral to that of nondegenerate plasma. Notice that for Coulomb scattering we have implemented a cutoff scheme based on minimal scattering angle \cite{2009PhRvD..79d3008A,1988A&A...191..181H}.

We consider mildly relativistic plasma with
\begin{equation}
0.01\lesssim e \lesssim100,
\end{equation}
where $e$ is particle kinetic energy divided by electron rest energy, this range contains both relativistic and non-relativistic domains. The upper limit is chosen to avoid thermal production of other particles such as neutrinos and muons, while the lower limit is required to have sufficient pair density.

We introduce logarithmic energy grid with $a_{max}=40$ nodes for all calculations and different homogeneous grids for angular variables, $\phi$-grid is 2 time denser then $\mu$-grid (typically $\mu$-grid contains $j_{max}=64$ nodes). To compare results with known analytical expressions we use definition of angle-averaged reaction rate per pair of particles 
\begin{equation}\label{sv1}
\overline{v\sigma}(e_1,e_2)=\int_{-1}^{+1} \frac{d\mu_{2}}2 \int_{\vec{p}_3,\vec{p}_4} v d\sigma,
\end{equation}
and angle-averaged emissivity per pair of particles 
\begin{equation}\label{sv2}
\overline{v\frac{d\sigma}{de_3}}(e_1,e_2,e_3)=\int_{-1}^{+1} \frac{d\mu_{2}}2 \int_{\vec{p}_3,\vec{p}_4} v \frac{d\sigma}{de_3},
\end{equation} introduced by Svensson~\cite{1982ApJ...258..321S}, where $d\sigma$ is given by standard definition \eqref{dsigma} and we have used spherical symmetry as described before Eq.~\eqref{stfI1}. We use Coppi \& Blandford \cite{1990MNRAS.245..453C} analytical expressions (2.3), (3.2), (4.3) for $\overline{v\sigma}$, which corresponds to $R_{ab}$. Svensson \cite{1982ApJ...258..321S} formula (55), Peer \& Waxman \cite{2005ApJ...628..857P} formulae (19, 28) are used for quantity $\overline{v\frac{d\sigma}{de}}$, which corresponds to $P_{abc}/\Delta e_a$.

To compare numerical results with analytical ones, we introduce the following quantity for each process
\begin{equation}
Q=\frac{1}{a_\text{max}^2}\sum_{a,b} |R_{ab}/\overline{v\sigma}(e_a,e_b)-1|,
\end{equation}
expressing average relative deviation of numerical results from analytical ones for all energy grid nodes. Table \ref{qtab} presents values of $Q$ for selected number of angular grid nodes. It is evident that the relative error decreases with increasing of number of angular grid nodes reaching about 1~\% with 128 nodes. This demonstrates convergence of numerical results to the corresponding analytical ones.

\begin{table}[tbp]
\caption{Values of $Q$ for selected number of angular grid nodes ($a_{max}=40, k_{max}=2j_{max}$). \label{qtab}} \center
\begin{tabular}[c]{|l|l|l|l|l|}
\hline
Process/$j_{max}$ & 16 & 32 & 64 & 128 \\
\hline
CS & 0.0855 & 0.0403 & 0.0207 & 0.0145\\
\hline
PA & 0.0231 & 0.00693 & 0.00313 & 0.00138\\
\hline
PC & 0.146 & 0.0657 & 0.0303 & 0.0116\\
\hline
\end{tabular}
\end{table}

Below we present some representative plots for the reaction rates of all reactions together with analytical curves (where they are known). Energy is measured in electron rest energy units. Presented results reproduce both nonrelativistic and relativistic energy cases. All computations were carried on Intel Core i3-7100 CPU @3.90 GHz processor using one processor core. The code is written in C and compiled in Windows 7 environment with Microsoft Visual Studio 2015 in fully optimized x64 mode. Computation time of initial angular integration of collision integrals for each reaction from Table~\ref{2pptable} is shown in Table~\ref{timing}. It shows even lower than expected $O(j_{max}^3)$ behaviour due to kinematic cuts on the phase space of reactions.

\begin{table}[tbp]
\caption{CPU time (in seconds) of each reaction initial angular integration for selected number of angular grid nodes ($a_{max}=40, k_{max}=2j_{max}$), and its exponent of computational cost $O(j_{max}^n)$. \label{timing}} \center
\begin{tabular}[c]{|l|l|l|l|l||l|}
\hline
Process/$j_{max}$ & 16 & 32 & 64 & 128 & n \\
\hline
%CS & 2.215 s & 14.476 s & 113.241 s & 590.089 s \\
CS & 2.215 & 14.48 & 113.2 & 590.1 & 2.7 \\
\hline
%PA & 2.106 s & 14.726 s & 100.168 s & 543.086 s \\
PA & 2.106 & 14.73 & 100.2 & 543.1 & 2.7 \\
\hline
%PC & 0.531 s & 3.619 s & 28.813 s & 223.206 s \\
PC & 0.531 & 3.619 & 28.82 & 223.2 & 2.9 \\
\hline
%MS & 2.418 s & 16.863 s & 130.526 s & 1030.277 s \\
MS & 2.418 & 16.87 & 130.5 & 1030 & 2.9 \\
\hline
%BS & 3.354 s & 22.744 s & 178.558 s & 1112.583 s \\
BS & 3.354 & 22.74 & 178.6 & 1113 & 2.8 \\
\hline
\end{tabular}
\end{table}

\begin{figure}[tbp]
\centering
\includegraphics[width=70mm]{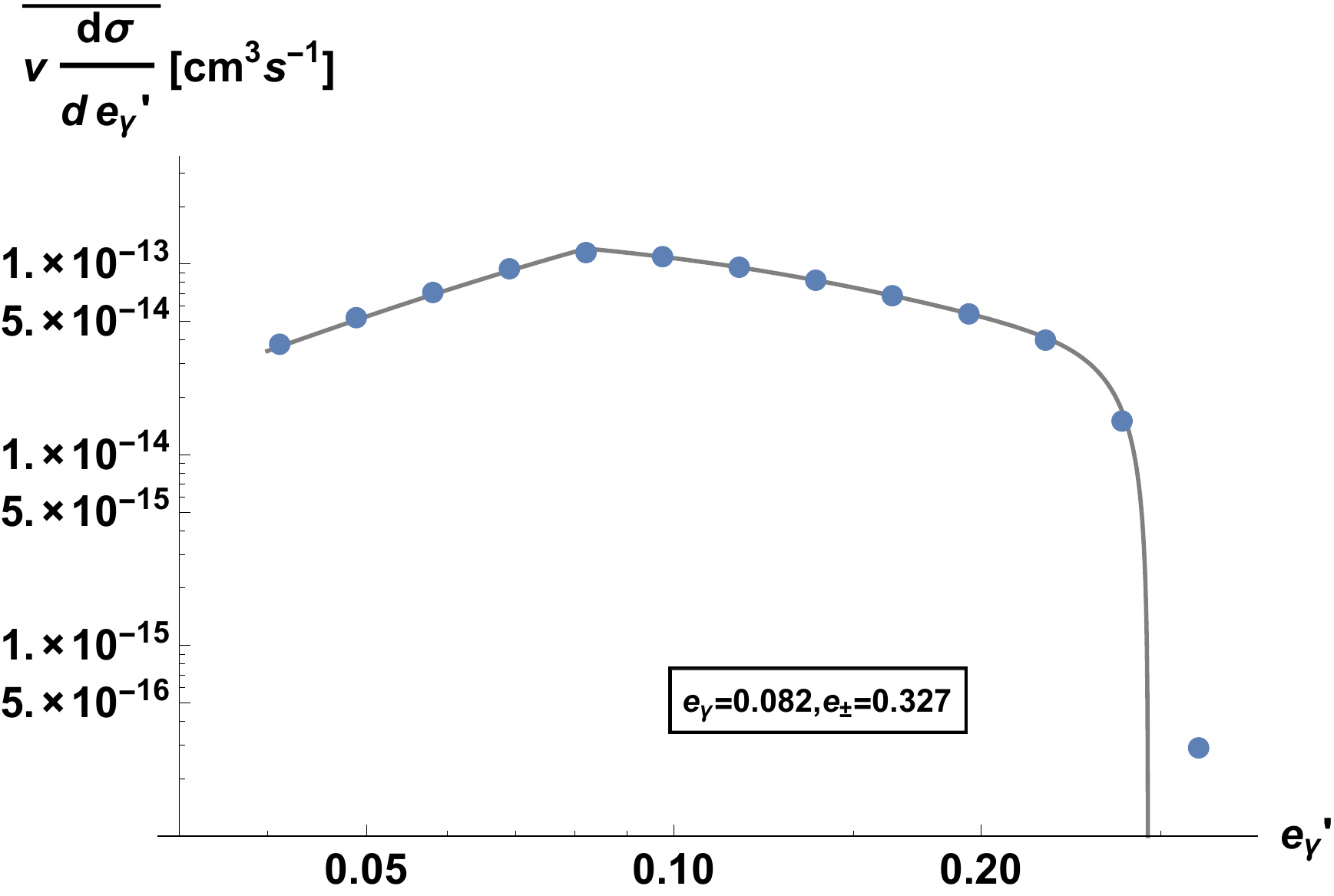}
\caption{Compton scattered-photon distribution $\overline{v\frac{d\sigma}{de_\gamma'}}(e_\gamma,e_{\pm},e_\gamma ').$}\label{CS_D}
\end{figure}

\begin{figure}[tbp]
\centering
\includegraphics[width=70mm]{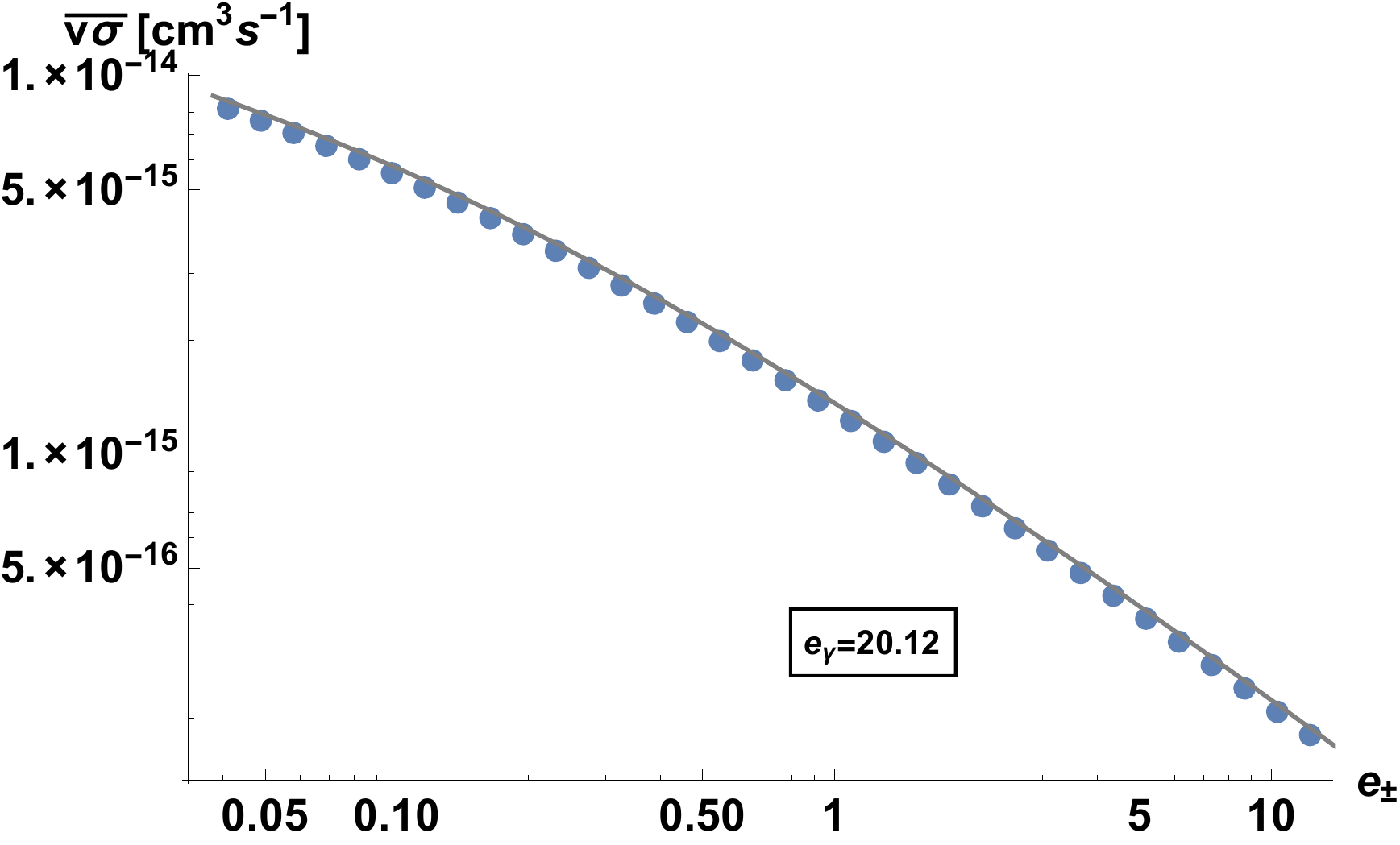}
\caption{Compton scattering rate $\overline{v\sigma}(e_\pm,e_\gamma)$.}\label{CS_I}
\end{figure}

Compton scattering presents well-known challenge for numerical treatment as all the analytical formulas for scattering rate behave badly numerically in different parameter areas, see e.g. \cite{2005ApJ...628..857P, 2009A&A...506..589B}. We easily bypass this difficulty as we numerically integrate well-behaved differential cross-section, as one can see for non-relativistic regime in Fig.~\ref{CS_D} and for relativistic regime in Fig.~\ref{CS_I}. Figure~\ref{CS_D} presents analytic photon spectrum for the reaction $\gamma+e^{\pm}\rightarrow\gamma'+e^{\pm}{}'$ as solid line and our numerical results shown by dots. Overall there is good agreement between numerical and analytical results. Small deviations in high-energy of the spectrum arise from leakage of the particles to kinematically forbidden area at the boundary of energy zones. Due to particle splitting (between cell interpolation) some final paricles would be placed on a grid node, that is kinematically forbidden, and it is indeed the case of Fig.~\ref{CS_D}. To show this effect we enlarge the plot  range especially on this figure. On the other spectrum figures these points appear to be outside the presented plot range. There the maximum allowed photon energy is $e_{max}=0.291$, but we have particles of energies from 0.275 up to $e_{max}$ that are splitted between energy zones of 0.275 and 0.327 -- the second is kinematically forbidden.
Figure~\ref{CS_I} shows the total reaction rate of the same process. Again there is good agreement between numerical and analytical results. Small discrepancy arises from truncation of reactions where final particles get out of the grid to higher or lower energies. As a result numerical reaction rates are systematically lower than analytic ones.

\begin{figure}[tbp]
\centering
\includegraphics[width=70mm]{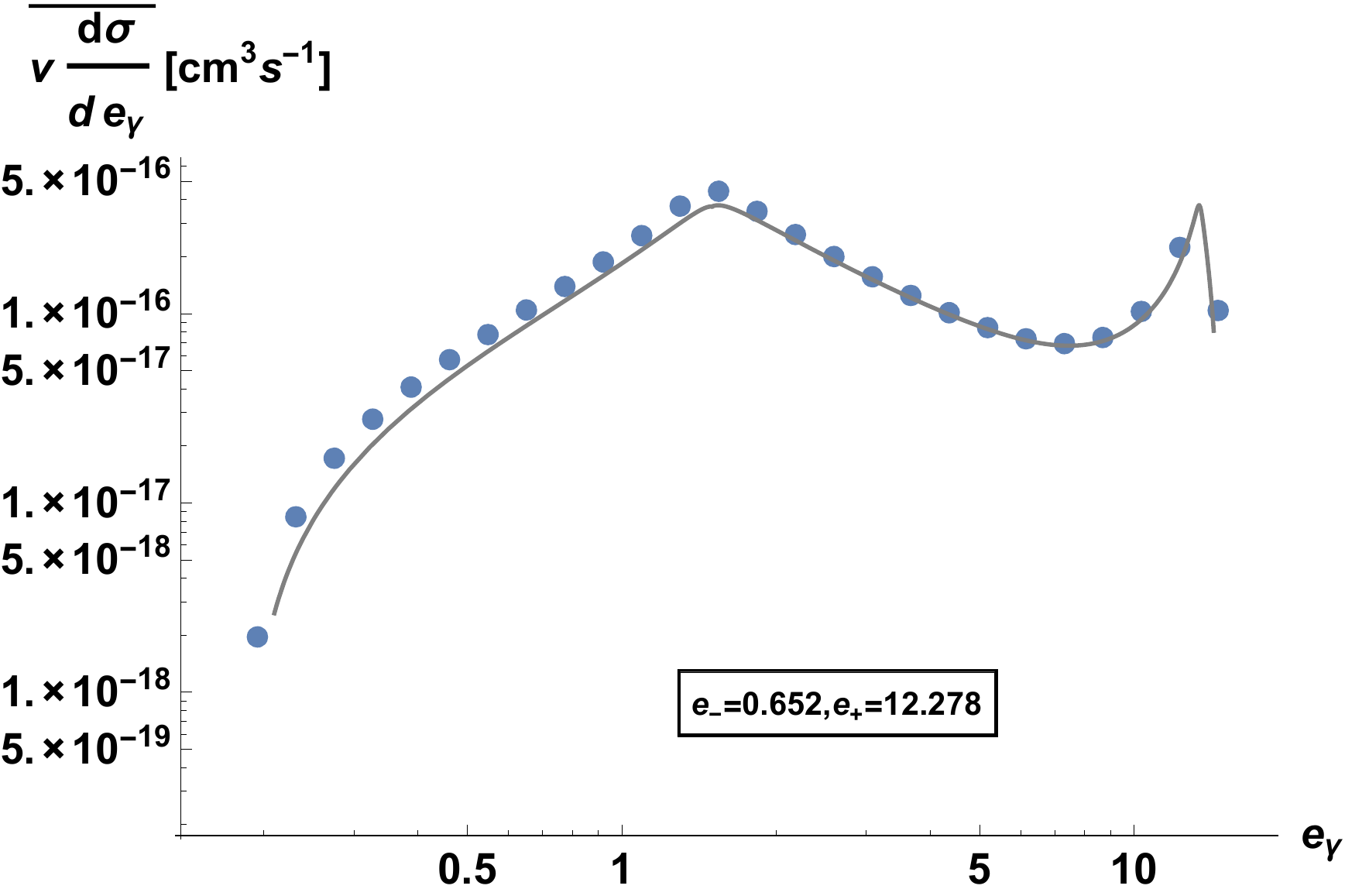}
\caption{Distribution for pair annihilation $\overline{v\frac{d\sigma}{de_\gamma}}(e_\gamma,e_{+},e_{-}).$}\label{AN_D}
\end{figure}

\begin{figure}[tbp]
\centering
\includegraphics[width=70mm]{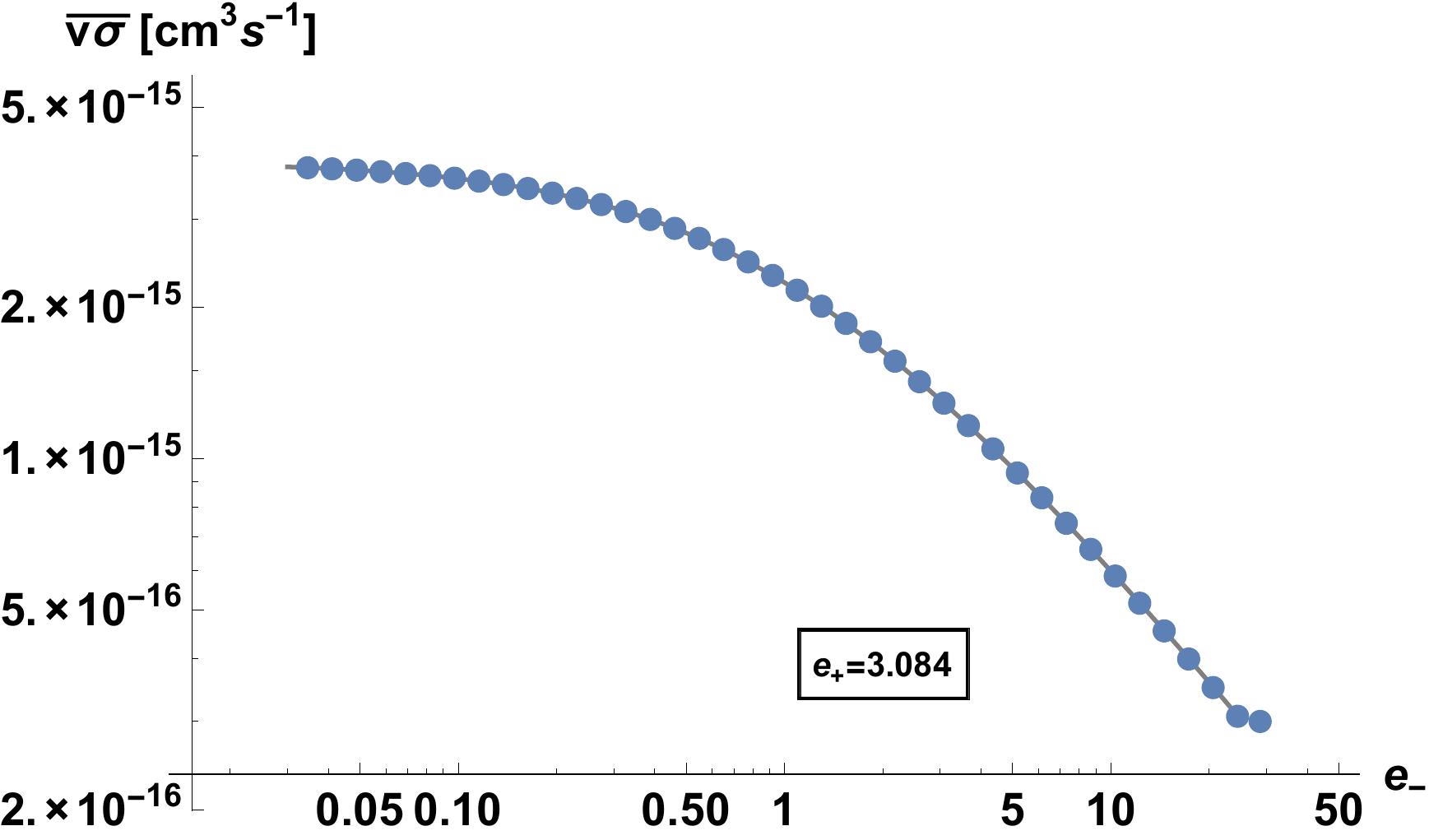}
\caption{Pair annihilation rate $\overline{v\sigma}(e_{+},e_{-}).$}\label{AN_I}
\end{figure}

Annihilation photon spectrum for reaction $e^{+}+e^{-}\rightarrow\gamma+\gamma'$ is illustrated in Fig.~\ref{AN_D} and total reaction rate in this process in Fig.~\ref{AN_I}. Figure~\ref{AN_D} shows that the method is able to accurately reproduce the spectrum of annihilation photons in the range of more than two orders of magnitude. Reaction truncation errors, hardly seen at Fig.~\ref{AN_I}, are much lower for annihilation as low-energy photons are rare in this process.

\begin{figure}[tbp]
\centering
\includegraphics[width=70mm]{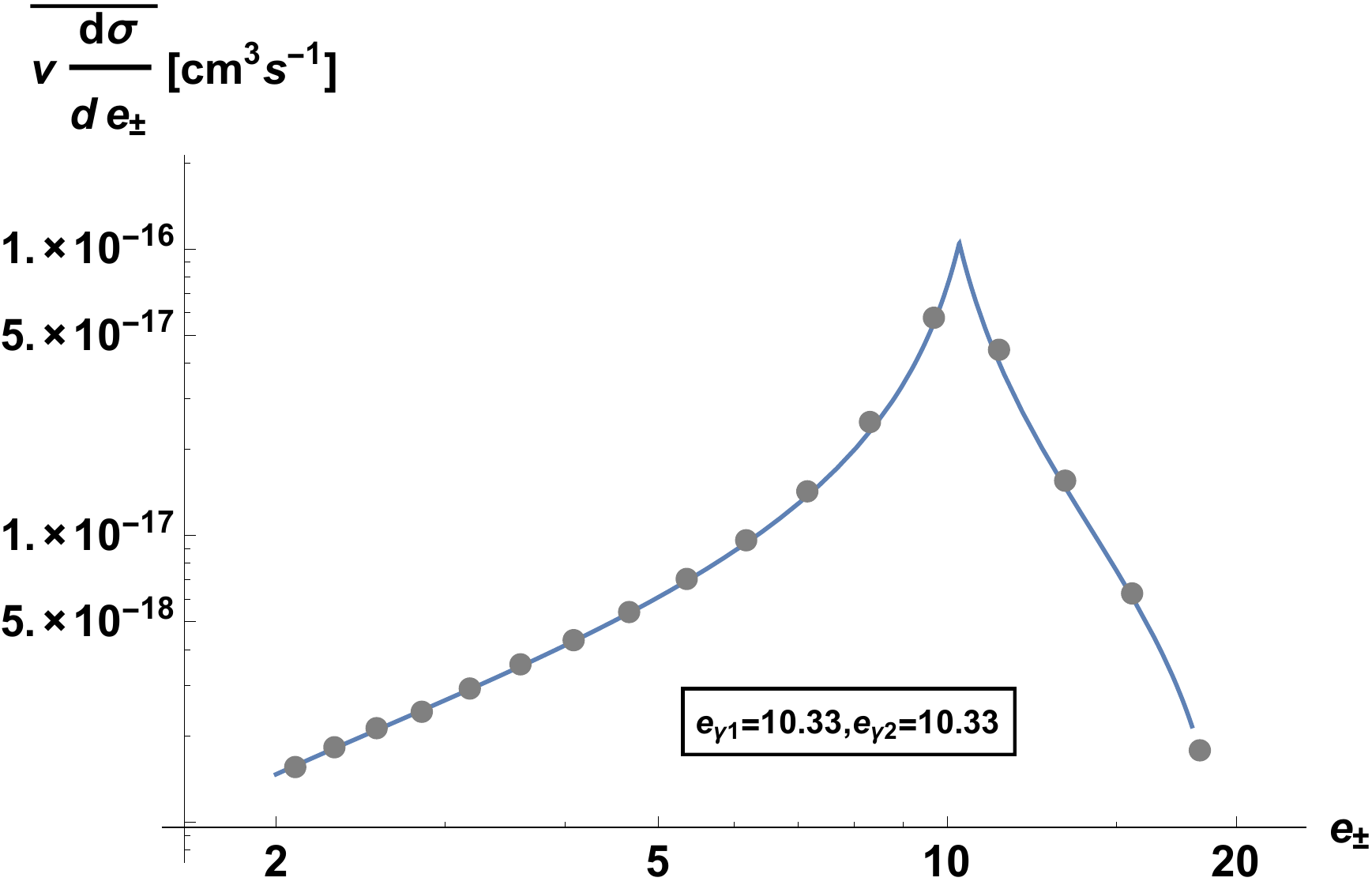}
\caption{Distribution for pair production $\overline{v\frac{d\sigma}{de_\pm}}(e_{\gamma_1},e_{\gamma_2},e_{\pm}).$}\label{PC_D}
\end{figure}

\begin{figure}[tbp]
\centering
\includegraphics[width=70mm]{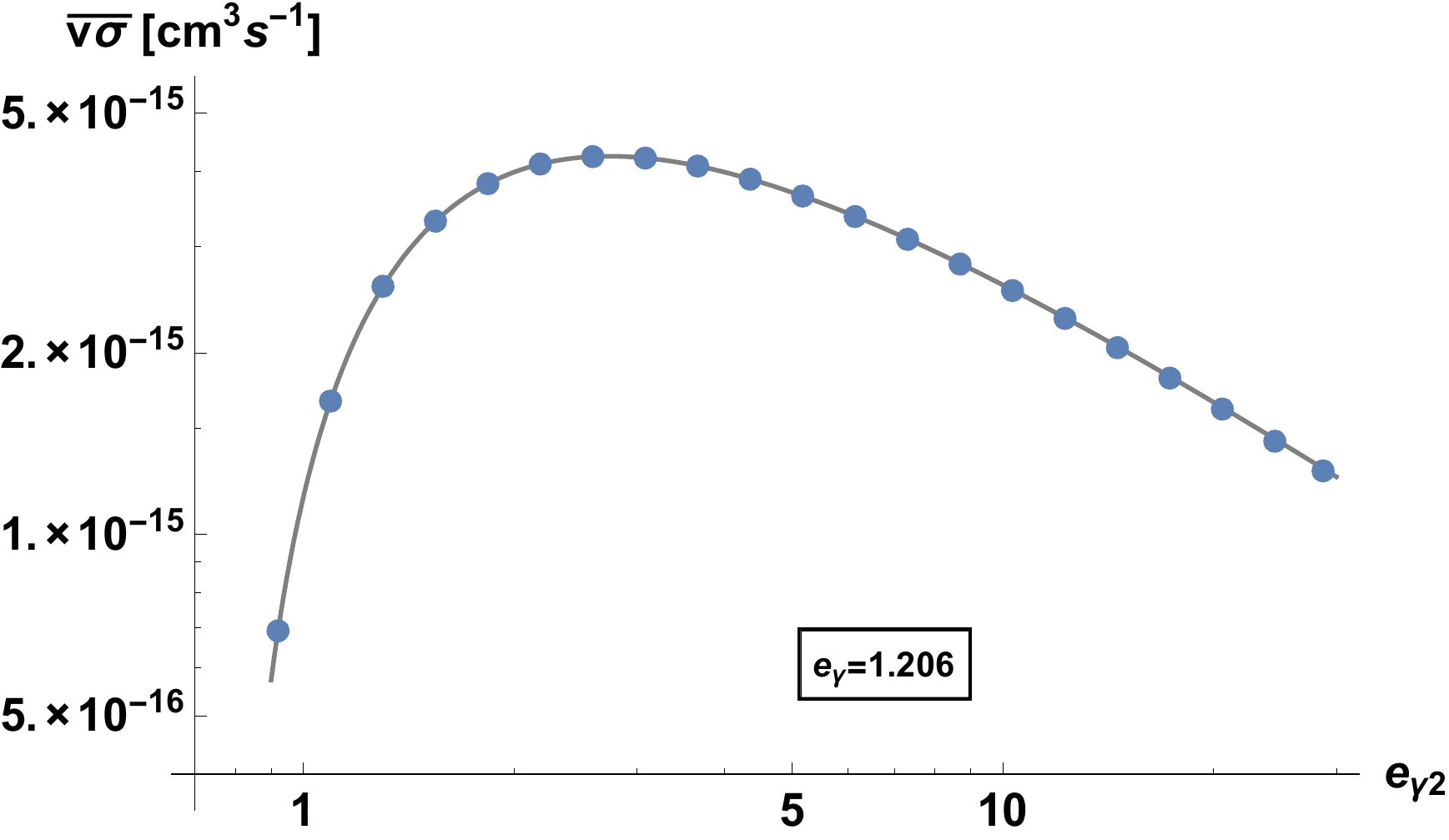}
\caption{Pair production rate $\overline{v\sigma}(e_{\gamma_1},e_{\gamma_2}).$}\label{PC_I}
\end{figure}

Balance between pair creation and annihilation represent an independent test for the numerical scheme, as it is not automatically satisfied due to different numerical treatment of incoming and outgoing particles in the reactions. Pair creation spectra for reaction $\gamma_1+\gamma_2\rightarrow e^{+}+e^{-}$ are reproduced well, see Fig.~\ref{PC_D}, as well as total reaction rates, see Fig.~\ref{PC_I}. Numerical balance can be checked by the form of particle distributions in numerical equilibrium, that was verified to be within 5~\% of corresponding Boltzmann distributions.

\begin{figure}[tbp]
\centering
\includegraphics[width=70mm]{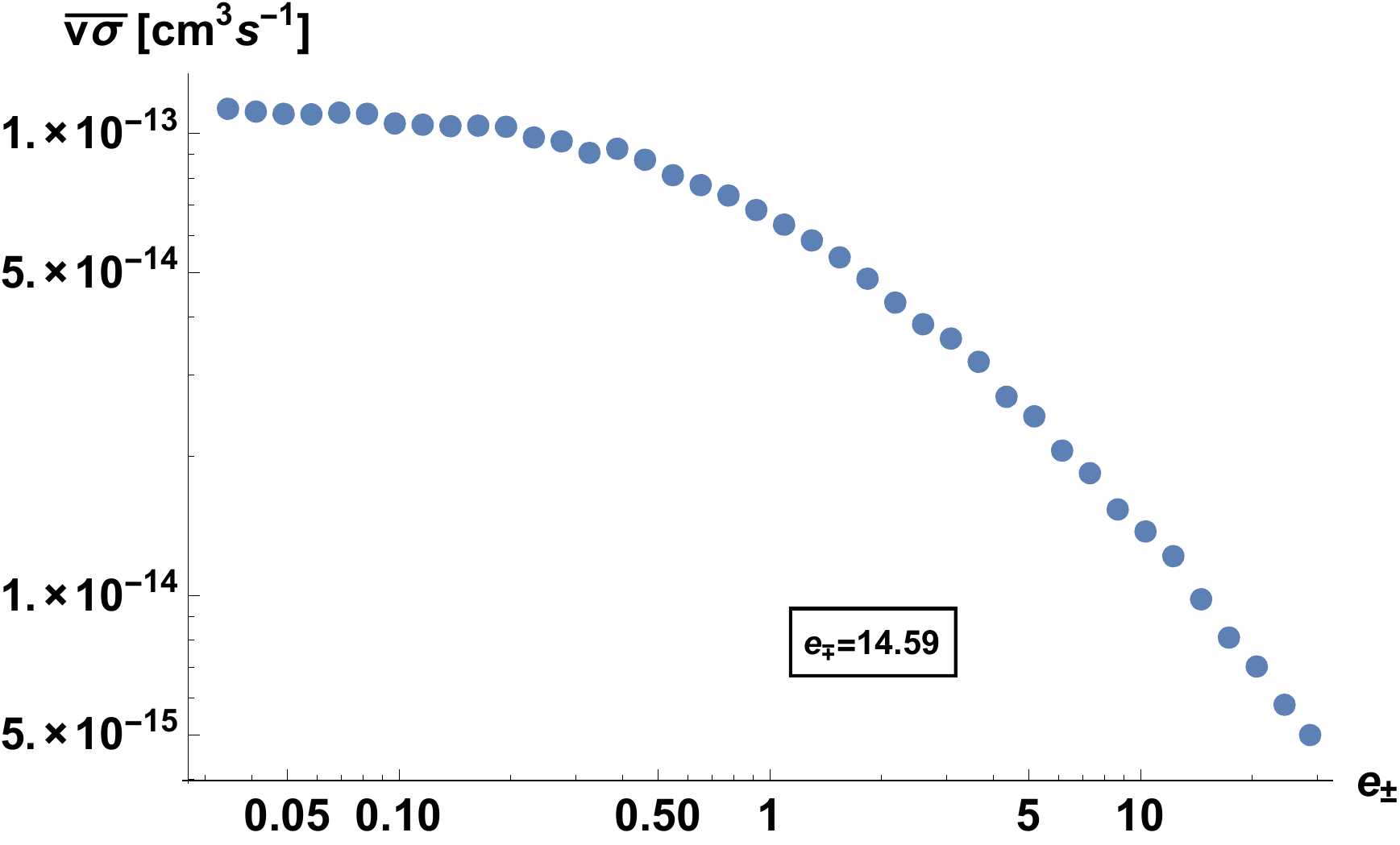}
\caption{Bhabha scattering rate $\overline{v\sigma}(e_+,e_-).$}\label{BS_I}
\end{figure}

\begin{figure}[tbp]
\centering
\includegraphics[width=70mm]{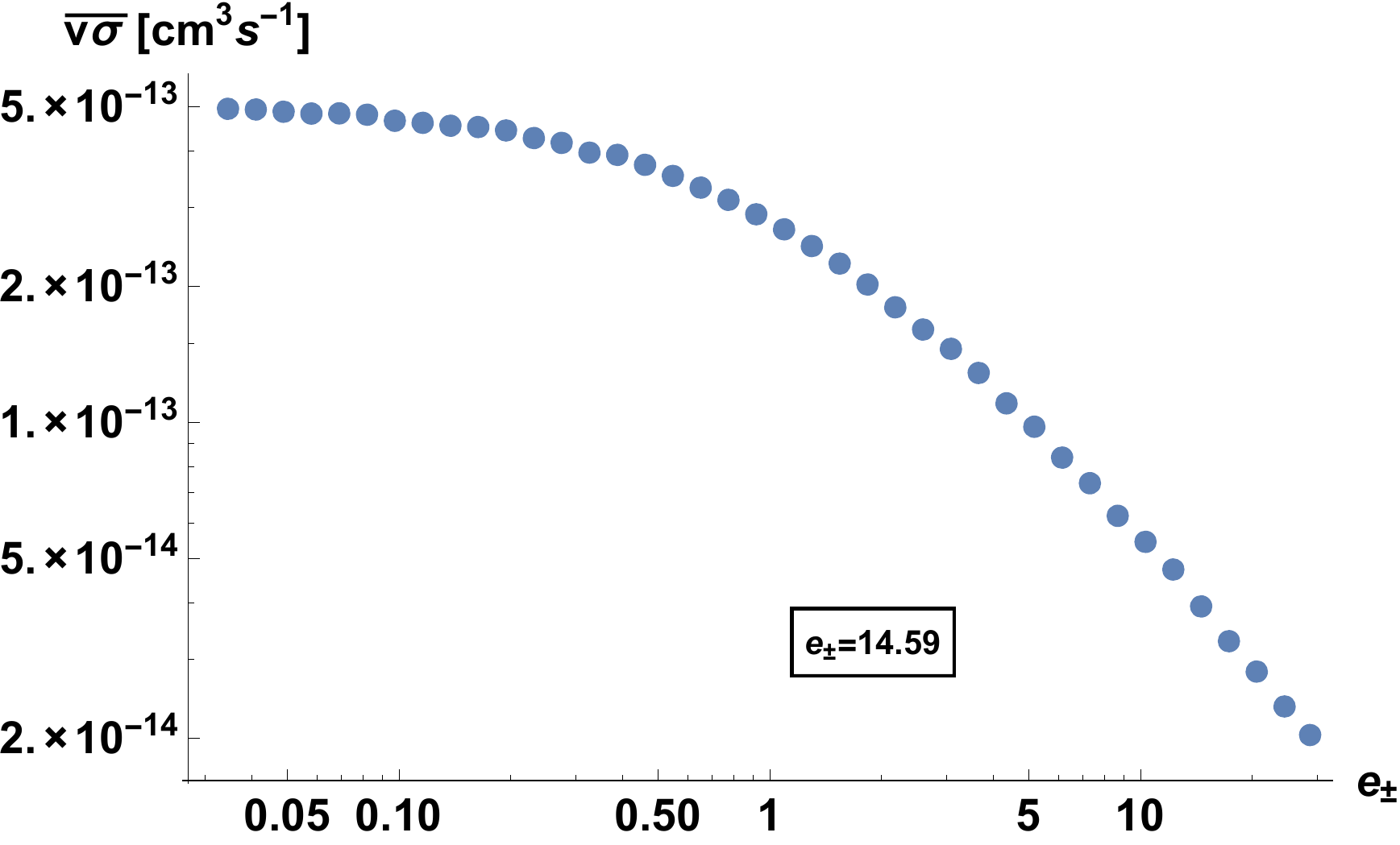}
\caption{M{\o}ller scattering rate $\overline{v\sigma}(e_{\pm},e_{\pm}).$\label{MS_I}}
\end{figure}

For completeness we present also the results for M{\o}ller and Bhabha scattering, they show that these processes are indeed dominant for electrons and positrons in relativistic plasma, compare Figs.~\ref{BS_I}, \ref{MS_I} with Figs.~\ref{CS_I}, \ref{AN_I}, \ref{PC_I}.

%\begin{figure}[tbp]
%\centering
%\includegraphics[width=80mm]{figsp1}
%\caption{\textcolor{red}{Initial photon spectrum. $t_{initial}=10^{-40} s$.}}\label{initspg}
%\end{figure}

\begin{figure}[ptb!]
\centering
\begin{tabular}{cc}
\includegraphics[width=70mm]{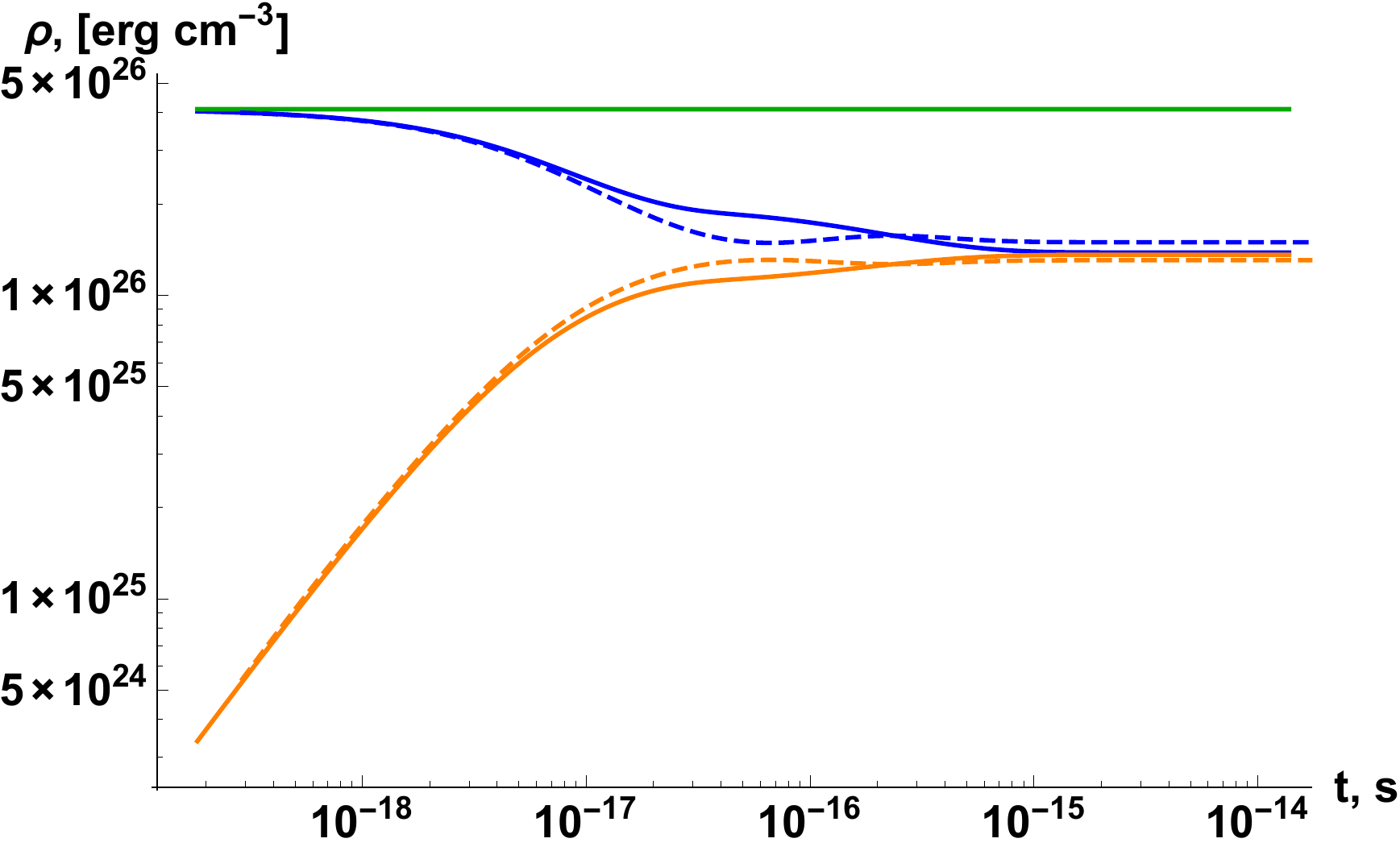}
\end{tabular}
\caption{Time evolution of energy density in components of electron-positron-photon plasma: photon energy density (blue), electron/positron energy density (orange), total energy density (green). Solid lines correspond to Boltzmann case, dashed lines correspond to Uehling-Uhlenbeck case.}\label{evolro}
\end{figure}
\begin{figure}[ptb!]
\centering
\begin{tabular}{cc}
\includegraphics[width=70mm]{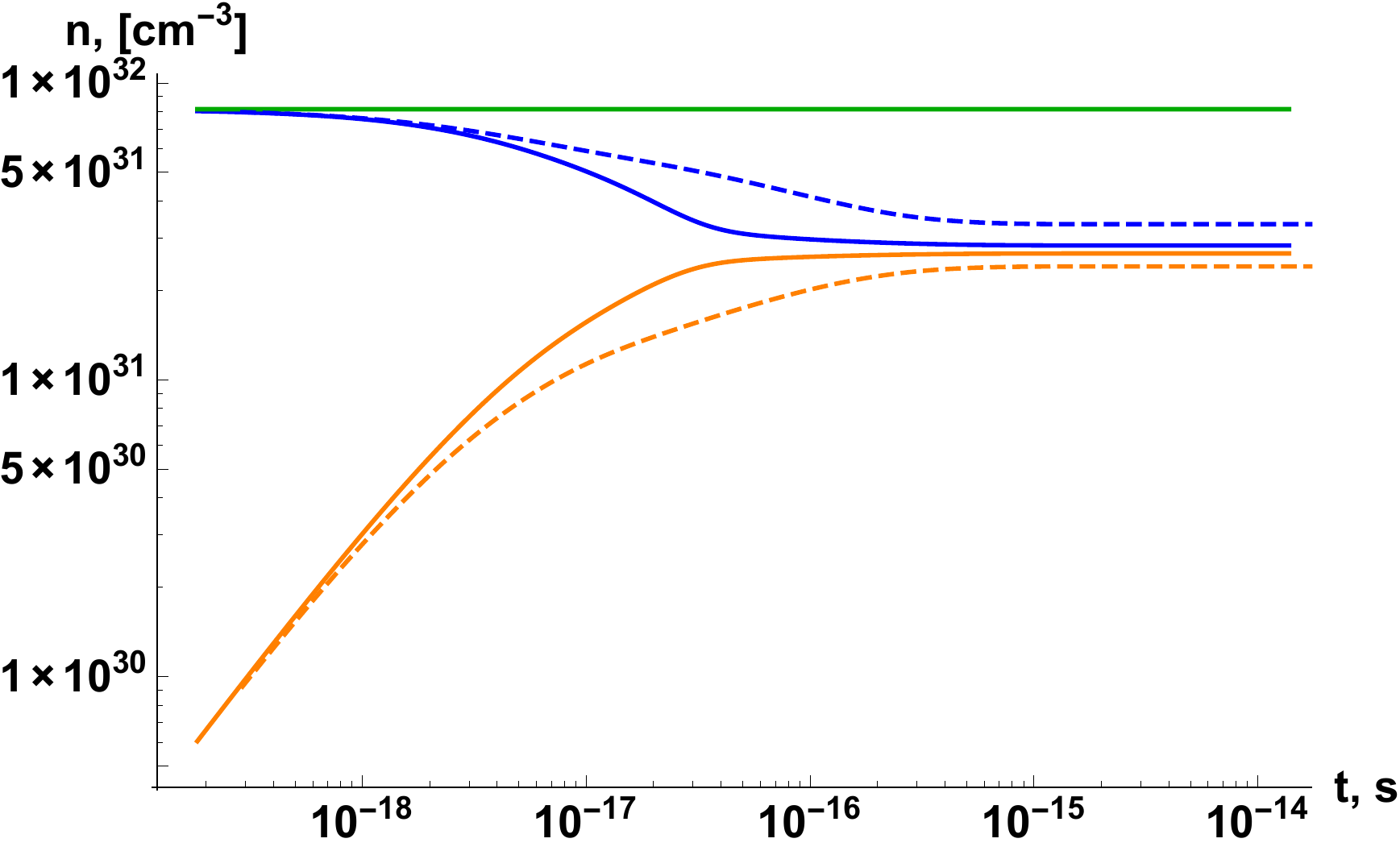}
\end{tabular}
\caption{Time evolution of number density in components of electron-positron-photon plasma: photon concentration (blue), electron/positron concentration (orange), total concentration (green). Solid lines correspond to Boltzmann case, dashed lines correspond to Uehling-Uhlenbeck case. Note the difference in the final pair and photon density due to rest mass of elecron/positron (in both cases) and difference in statistics (for U-U case).}\label{evoln}
\end{figure}

\begin{figure}[ptb!]
\centering
\begin{tabular}{cc}
\includegraphics[width=70mm]{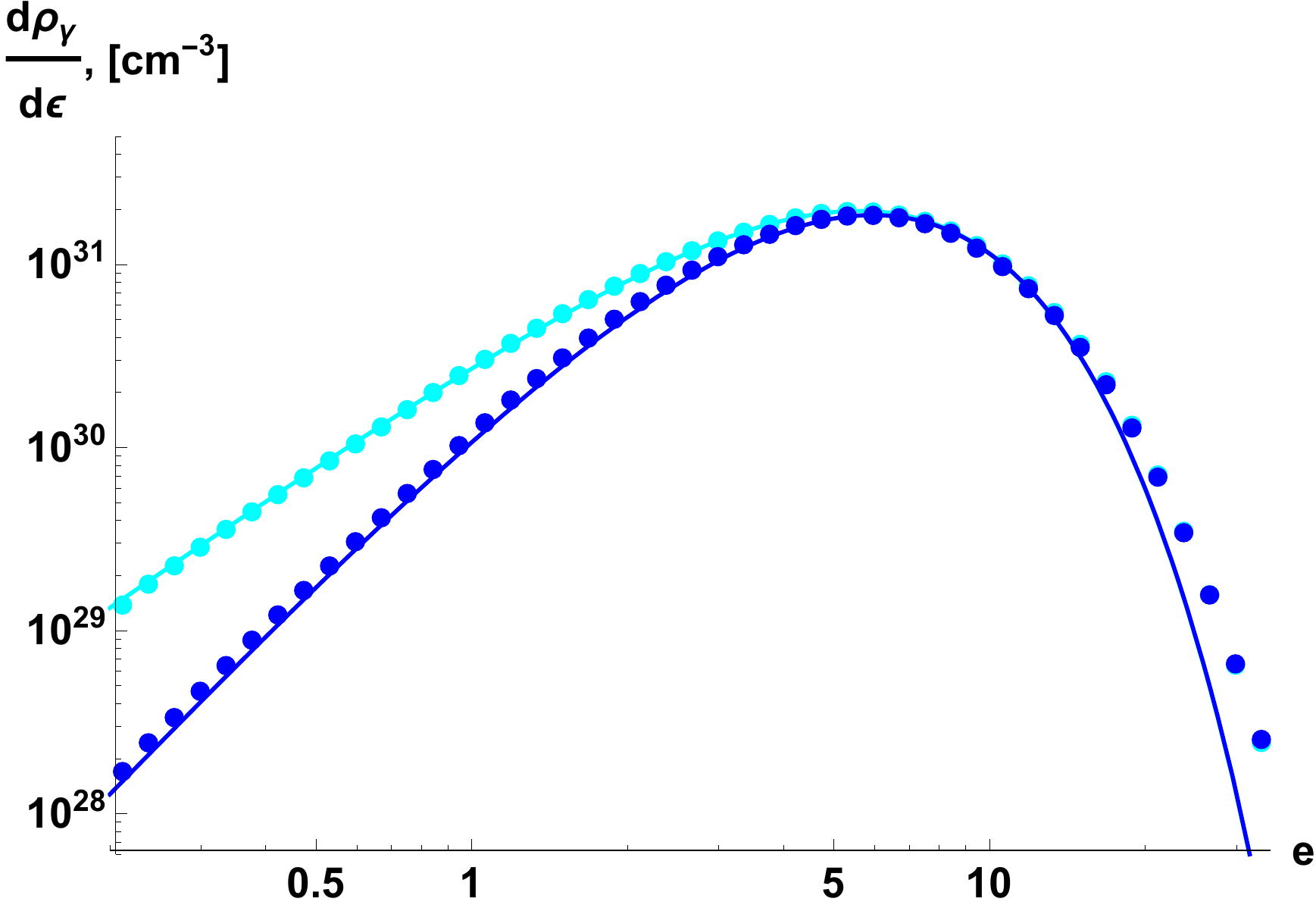} &\includegraphics[width=70mm]{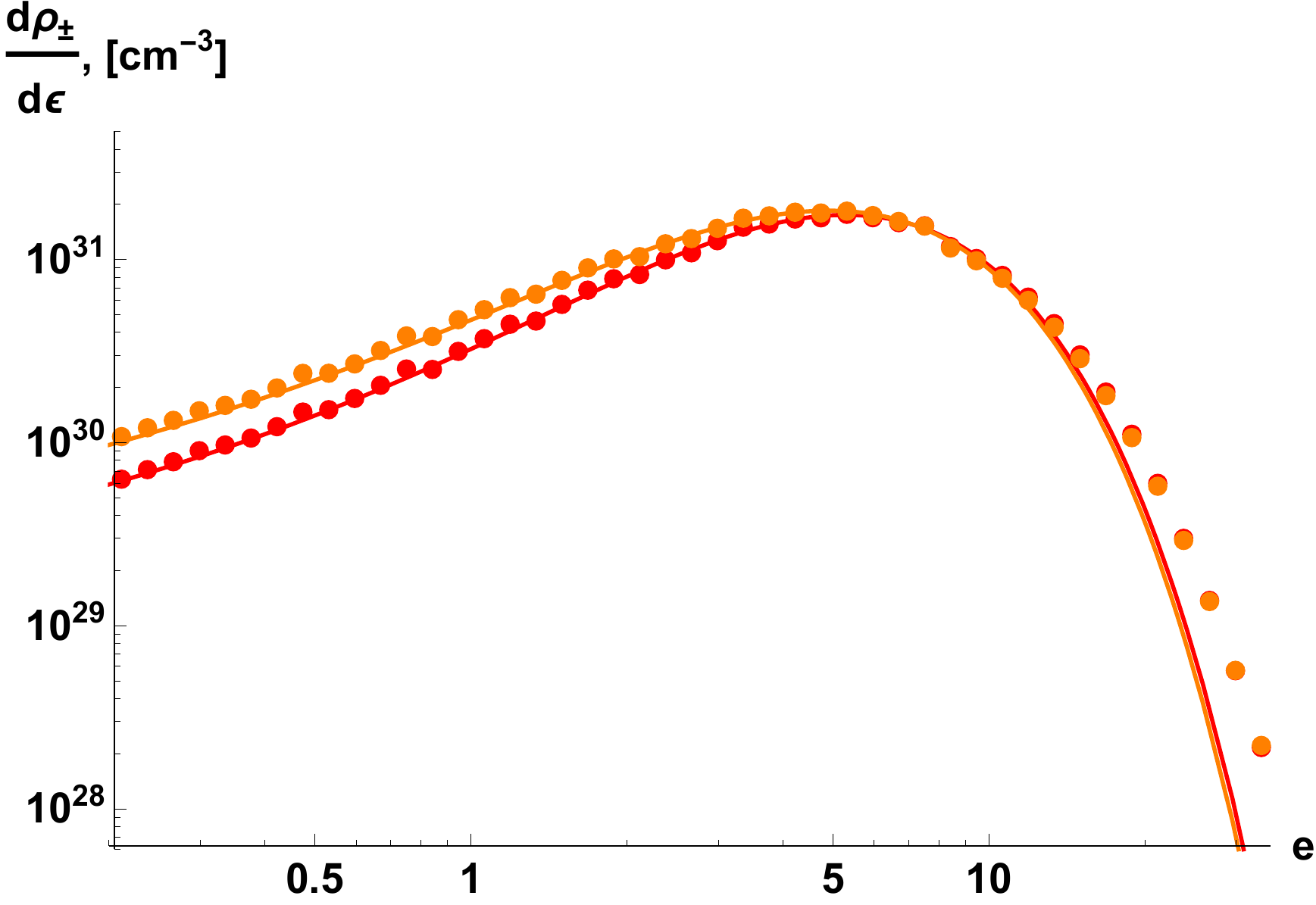}
\end{tabular}
\caption{Final energy spectra at $t_{final}=10^{-14}$~s. Solid lines are equilibrium Boltzmann and Bose-Einstein/Fermi-Dirac fits of numerical results: photon Boltzmann energy spectrum (blue), photon Bose-Einstein energy spectrum (cyan), pairs Boltzmann energy spectrum (orange), pairs Fermi-Dirac energy spectrum (red).}\label{evolge}
\end{figure}
{Finally, we present a time evolution of energy density and concentration to demonstrate the difference between the classical Boltzmann and U-U equations. Both systems \eqref{BoltzmannFinal} and \eqref{BoltzmannFinal2} were solved numerically with the same initial conditions under $a_{max}=60, j_{max}=64, k_{max}=2j_{max}$ and using Gear's method for resulting stiff ODE system \cite{1976oup..book.....H}.

The energy spectrum $d\rho/d\varepsilon$ is shown instead of the distribution function $f$, that are related by $d\rho/d\varepsilon=4\pi |\mathbf{p}|\varepsilon^2 c^{-2}f$. We chose an initial state without electrons and positrons but with photons only, initial spectrum has a power-law shape $d\rho/d\varepsilon=a (\varepsilon/\varepsilon_0)^b$, with $a=3.63\times10^{28}\text{ cm}^{-3}$ and $b=-0.438$, $\varepsilon_0=1$~erg, between $e=0.157$ and $e=157$. The initial spectrum corresponds to a total energy density $\rho=4.10\times 10^{26}\text{ erg cm}^{-3}$ and a total number density of particles $n=8.15\times 10^{31}\text{ cm}^{-3}$. In general, initial spectrum can have an arbitrary shape and thermalization process transforms it to an equilibrium form. Fig.~\ref{evolge} represents energy spectra at final equilibrium state. They attain corresponding shapes of Boltzmann and Bose-Einstein/Fermi-Dirac with some deviations in high-energy tails that are attributed to reaction truncation errors described before. We note that total energy and number densities do not change in time due to particle splitting applied, this feature does not depend on a form of a system of equations or a type of numerical ODE solver.} \\

\newpage
\section{Conclusions}\label{ch6}

In this paper, we propose a new numerical method to accurately calculate Uehling–Uhlenbeck collision integral for two-particle interactions in relativistic plasma. Exact energy and particle number conservation laws are achieved by using interpolation scheme \eqref{csol}. After calculation of collision integral discretized Uehling–Uhlenbeck equations transforms into system of ODEs, which can be treated by various methods suitable to solve stiff ODEs. The method admits parallelization on GPU/CPU. Improvement in computation time with respect to previous work is achieved. Our reaction-oriented approach can be easily applied to any other types of particles and any other binary interactions, for instanse, weak interactions of neutrinos or electromagnetic ones of protons.
Generalization of the proposed method for triple interactions is straightforward.

Our results show that reaction rates in relativistic plasma are well reproduced with moderate number of grid nodes in energy and angles (see Figures and Table \ref{2pptable}) both for non-relativistic and relativistic particle energies. This allows development of an efficient method of solution for relativistic Uehling–Uhlenbeck equation. 

%Our next goal is an inclusion of triple interactions into U-U collision integral. This task takes place in connection with the thermalization in relativistic plasma. It was stated in \cite{2009PhRvD..79d3008A} that account of binary reactions in relativistic plasma does not allow to reach thermodynamic equilibrium state. The necessary condition of thermodynamic equilibrium state in relativistic plasma is detailed balance at least in three-particle interactions. Sufficient condition to reach thermodynamic equilibrium state is a full balance between all direct and inverse reactions. So that, in principal, one has to consider not only triple interaction but also interactions of higher order in loop expansion. The timescale of reaching thermodynamic equalibrium is defined by the most slow nonbalanced reaction.
\section{Acknowledgements}
We thank anonymous referees for their remarks which improved the presentation of our results.
\newpage

\bibliographystyle{unsrt}
\bibliography{litera}

\begin{thebibliography}{10}

\bibitem{weinberg2008cosmology}
S.~Weinberg.
\newblock {\em Cosmology}.
\newblock OUP Oxford, 2008.

\bibitem{1999PhR...314..575P}
T.~{Piran}.
\newblock {Gamma-ray bursts and the fireball model}.
\newblock {\em Physics Reports}, 314:575--667, June 1999.

\bibitem{2010PhR...487....1R}
R.~{Ruffini}, G.~{Vereshchagin}, and S.-S. {Xue}.
\newblock {Electron-positron pairs in physics and astrophysics: From heavy
  nuclei to black holes}.
\newblock {\em Physics Reports}, 487:1--140, February 2010.

\bibitem{2015PhR...561....1K}
P.~{Kumar} and B.~{Zhang}.
\newblock {The physics of gamma-ray bursts \& relativistic jets}.
\newblock {\em Physics Reports}, 561:1--109, February 2015.

\bibitem{2012AAT...27..557A}
R.~{Antonucci}.
\newblock {A panchromatic review of thermal and nonthermal active galactic
  nuclei}.
\newblock {\em Astronomical and Astrophysical Transactions}, 27:557--602, 2012.

\bibitem{blandford2013active}
P.R.D. Blandford, P.H. Netzer, P.L. Woltjer, T.J.L. Courvoisier, and P.M.
  Mayor.
\newblock {\em Active Galactic Nuclei}.
\newblock Saas-Fee Advanced Course. Springer Berlin Heidelberg, 2013.

\bibitem{2006ARA&A..44..323F}
G.~{Fabbiano}.
\newblock {Populations of X-Ray Sources in Galaxies}.
\newblock {\em Annual Review of Astronomy and Astrophysics}, 44:323--366,
  September 2006.

\bibitem{2015NatCo...6E6747S}
G.~{Sarri}, K.~{Poder}, J.~M. {Cole}, W.~{Schumaker}, A.~{di Piazza},
  B.~{Reville}, T.~{Dzelzainis}, D.~{Doria}, L.~A. {Gizzi}, G.~{Grittani},
  S.~{Kar}, C.~H. {Keitel}, K.~{Krushelnick}, S.~{Kuschel}, S.~P.~D. {Mangles},
  Z.~{Najmudin}, N.~{Shukla}, L.~O. {Silva}, D.~{Symes}, A.~G.~R. {Thomas},
  M.~{Vargas}, J.~{Vieira}, and M.~{Zepf}.
\newblock {Generation of neutral and high-density electron-positron pair
  plasmas in the laboratory}.
\newblock {\em Nature Communications}, 6:6747, April 2015.

\bibitem{0741-3335-53-1-015009}
R~Duclous, J~G Kirk, and A~R Bell.
\newblock Monte carlo calculations of pair production in high-intensity
  laser-plasma interactions.
\newblock {\em Plasma Physics and Controlled Fusion}, 53(1):015009, 2011.

\bibitem{1742-6596-454-1-012016}
Toseo Moritaka, Luca Baiotti, An~Lin, Li~Weiwu, Youichi Sakawa, Yasuhiro
  Kuramitsu, Taichi Morita, and Hideaki Takabe.
\newblock Plasma particle-in-cell simulations with qed reactions for pair
  production experiments using a high-z solid target.
\newblock {\em Journal of Physics: Conference Series}, 454(1):012016, 2013.

\bibitem{PhysRevLett.102.105001}
Hui Chen, Scott~C. Wilks, James~D. Bonlie, Edison~P. Liang, Jason Myatt,
  Dwight~F. Price, David~D. Meyerhofer, and Peter Beiersdorfer.
\newblock Relativistic positron creation using ultraintense short pulse lasers.
\newblock {\em Phys. Rev. Lett.}, 102:105001, Mar 2009.

\bibitem{vereshchagin2017relativistic}
G.V. Vereshchagin and A.G. Aksenov.
\newblock {\em Relativistic Kinetic Theory: With Applications in Astrophysics
  and Cosmology}.
\newblock Cambridge University Press, 2017.

\bibitem{cercignani2012relativistic}
C.~Cercignani and G.M. Kremer.
\newblock {\em The Relativistic Boltzmann Equation: Theory and Applications}.
\newblock Progress in Mathematical Physics. Birkh{\"a}user Basel, 2012.

\bibitem{groot1980relativistic}
S.R. Groot, W.A. Leeuwen, and C.G. Weert.
\newblock {\em Relativistic kinetic theory: principles and applications}.
\newblock North-Holland Pub. Co., 1980.

\bibitem{1990JMP....31..245B}
N.~{Bellomo} and S.~{Kawashima}.
\newblock {The discrete Boltzmann equation with multiple collisions: Global
  existence and stability for the initial value problem}.
\newblock {\em Journal of Mathematical Physics}, 31:245--253, January 1990.

\bibitem{1991RvMaP...3..137B}
N.~{Bellomo} and T.~{Gustafsson}.
\newblock {The Discrete Boltzmann Equation:. a Review of the Mathematical
  Aspects of the Initial and Initial-Boundary Value Problems}.
\newblock {\em Reviews in Mathematical Physics}, 3:137--162, 1991.

\bibitem{dimarco_pareschi_2014}
G.~Dimarco and L.~Pareschi.
\newblock Numerical methods for kinetic equations.
\newblock {\em Acta Numerica}, 23:369–520, 2014.

\bibitem{2006MaCom..75.1833M}
C.~{Mouhot} and L.~{Pareschi}.
\newblock {Fast algorithms for computing the Boltzmann collision operator}.
\newblock {\em Mathematics of Computation}, 75:1833--1852, December 2006.

\bibitem{DIMARCO2017}
Giacomo Dimarco, Raphaël Loubère, Jacek Narski, and Thomas Rey.
\newblock An efficient numerical method for solving the boltzmann equation in
  multidimensions.
\newblock {\em Journal of Computational Physics}, 2017.

\bibitem{WU201327}
Lei Wu, Craig White, Thomas~J. Scanlon, Jason~M. Reese, and Yonghao Zhang.
\newblock Deterministic numerical solutions of the boltzmann equation using the
  fast spectral method.
\newblock {\em Journal of Computational Physics}, 250(Supplement C):27 -- 52,
  2013.

\bibitem{SHERLOCK20082286}
M.~Sherlock.
\newblock A monte-carlo method for coulomb collisions in hybrid plasma models.
\newblock {\em Journal of Computational Physics}, 227(4):2286 -- 2292, 2008.

\bibitem{HUTHMACHER2016535}
Klaus Huthmacher, Andreas~K. Molberg, Bärbel Rethfeld, and Jeremy~R. Gulley.
\newblock A split-step method to include electron–electron collisions via
  monte carlo in multiple rate equation simulations.
\newblock {\em Journal of Computational Physics}, 322(Supplement C):535 -- 546,
  2016.

\bibitem{TURRELL2015144}
A.E. Turrell, M.~Sherlock, and S.J. Rose.
\newblock Self-consistent inclusion of classical large-angle coulomb collisions
  in plasma monte carlo simulations.
\newblock {\em Journal of Computational Physics}, 299(Supplement C):144 -- 155,
  2015.

\bibitem{BOBYLEV2013123}
A.V. Bobylev and I.F. Potapenko.
\newblock Monte carlo methods and their analysis for coulomb collisions in
  multicomponent plasmas.
\newblock {\em Journal of Computational Physics}, 246(Supplement C):123 -- 144,
  2013.

\bibitem{1981phki.book.....L}
E.~M. {Lifshitz} and L.~P. {Pitaevskii}.
\newblock {\em {Physical kinetics}}.
\newblock 1981.

\bibitem{2009PhRvD..79d3008A}
A.~G. {Aksenov}, R.~{Ruffini}, and G.~V. {Vereshchagin}.
\newblock {Thermalization of the mildly relativistic plasma}.
\newblock {\em Phys. Rev. D}, 79(4):043008, February 2009.

\bibitem{1934PhRv...46..917U}
E.~A. {Uehling}.
\newblock {Transport Phenomena in Einstein-Bose and Fermi-Dirac Gases. II}.
\newblock {\em Physical Review}, 46:917--929, November 1934.

\bibitem{1933PhRv...43..552U}
E.~A. {Uehling} and G.~E. {Uhlenbeck}.
\newblock {Transport Phenomena in Einstein-Bose and Fermi-Dirac Gases. I}.
\newblock {\em Physical Review}, 43:552--561, April 1933.

\bibitem{2017JCoPh.330.1010Y}
R.~{Yano}.
\newblock {Fast and accurate calculation of dilute quantum gas using
  Uehling-Uhlenbeck model equation}.
\newblock {\em Journal of Computational Physics}, 330:1010--1021, February
  2017.

\bibitem{Hu2015}
Jingwei Hu, Qin Li, and Lorenzo Pareschi.
\newblock Asymptotic-preserving exponential methods for the quantum boltzmann
  equation with high-order accuracy.
\newblock {\em Journal of Scientific Computing}, 62(2):555--574, Feb 2015.

\bibitem{huying12}
Jingwei Hu and Lexing Ying.
\newblock A fast spectral algorithm for the quantum boltzmann collision
  operator.
\newblock {\em Commun. Math. Sci.}, 10(3):989--999, 2012.

\bibitem{PhysRevE.68.056703}
Alejandro~L. Garcia and Wolfgang Wagner.
\newblock Direct simulation monte carlo method for the
  uehling-uhlenbeck-boltzmann equation.
\newblock {\em Phys. Rev. E}, 68:056703, Nov 2003.

\bibitem{2010arXiv1009.3352F}
F.~{Filbet}, J.~{Hu}, and S.~{Jin}.
\newblock {A Numerical Scheme for the Quantum Boltzmann Equation Efficient in
  the Fluid Regime}.
\newblock {\em ArXiv e-prints}, September 2010.

\bibitem{2004ApJ...609..363A}
A.~G. {Aksenov}, M.~{Milgrom}, and V.~V. {Usov}.
\newblock {Structure of Pair Winds from Compact Objects with Application to
  Emission from Hot Bare Strange Stars}.
\newblock {\em Astrophysical Journal}, 609:363--377, July 2004.

\bibitem{2007PhRvL..99l5003A}
A.~G. {Aksenov}, R.~{Ruffini}, and G.~V. {Vereshchagin}.
\newblock {Thermalization of Nonequilibrium Electron-Positron-Photon Plasmas}.
\newblock {\em Physical Review Letters}, 99(12):125003, September 2007.

\bibitem{2009AIPC.1111..344A}
A.~G. {Aksenov}, R.~{Ruffini}, and G.~V. {Vereshchagin}.
\newblock {Thermalization of pair plasma with proton loading}.
\newblock In G.~{Giobbi}, A.~{Tornambe}, G.~{Raimondo}, M.~{Limongi}, L.~A.
  {Antonelli}, N.~{Menci}, and E.~{Brocato}, editors, {\em American Institute
  of Physics Conference Series}, volume 1111 of {\em American Institute of
  Physics Conference Series}, pages 344--350, May 2009.

\bibitem{2010AIPC.1205...11A}
A.~G. {Aksenov}, R.~{Ruffini}, and G.~V. {Vereshchagin}.
\newblock {Kinetics of the mildly relativistic plasma and GRBs}.
\newblock In R.~{Ruffini} and G.~{Vereshchagin}, editors, {\em American
  Institute of Physics Conference Series}, volume 1205 of {\em American
  Institute of Physics Conference Series}, pages 11--16, March 2010.

\bibitem{2010PhRvE..81d6401A}
A.~G. {Aksenov}, R.~{Ruffini}, and G.~V. {Vereshchagin}.
\newblock {Pair plasma relaxation time scales}.
\newblock {\em Phys. Rev. E}, 81(4):046401, April 2010.

\bibitem{1973rela.conf....1E}
J.~{Ehlers}.
\newblock {Survey of general relativity theory}.
\newblock In {\em Relativity, Astrophysics and Cosmology}, pages 1--125, 1973.

\bibitem{2003spr..book.....G}
W.~{Greiner} and J.~{Reinhardt}.
\newblock {\em {Quantum Electrodynamics}}.
\newblock Berlin, Springer, 2003.

\bibitem{1982els..book.....B}
V.~B. {Berestetskii}, E.~M. {Lifshitz}, and V.~B. {Pitaevskii}.
\newblock {\em {Quantum Electrodynamics}}.
\newblock Elsevier, $2^\text{nd}$ edition, 1982.

\bibitem{BENEDETTI2013206}
A.~Benedetti, R.~Ruffini, and G.V. Vereshchagin.
\newblock Phase space evolution of pairs created in strong electric fields.
\newblock {\em Physics Letters A}, 377(3):206 -- 215, 2013.

\bibitem{ivanphd}
Siutsou I.
\newblock PhD thesis, University of Rome, Sapienza, 2013.

\bibitem{2015AIPC.1693g0007S}
I.~A. {Siutsou}, A.~G. {Aksenov}, and G.~V. {Vereshchagin}.
\newblock {On thermalization of electron-positron-photon plasma}.
\newblock In {\em American Institute of Physics Conference Series}, volume 1693
  of {\em American Institute of Physics Conference Series}, page 070007,
  December 2015.

\bibitem{1988A&A...191..181H}
E.~{Haug}.
\newblock {Energy loss and mean free path of electrons in a hot thermal
  plasma}.
\newblock {\em Astronomy and Astrophysics}, 191:181--185, February 1988.

\bibitem{1982ApJ...258..321S}
R.~{Svensson}.
\newblock {The pair annihilation process in relativistic plasmas}.
\newblock {\em Astrophysical Journal}, 258:321--334, July 1982.

\bibitem{1990MNRAS.245..453C}
P.~S. {Coppi} and R.~D. {Blandford}.
\newblock {Reaction rates and energy distributions for elementary processes in
  relativistic pair plasmas}.
\newblock {\em MNRAS}, 245:453--507, August 1990.

\bibitem{2005ApJ...628..857P}
A.~{Pe'er} and E.~{Waxman}.
\newblock {Time-dependent Numerical Model for the Emission of Radiation from
  Relativistic Plasma}.
\newblock {\em Astrophysical Journal}, 628:857--866, August 2005.

\bibitem{2009A&A...506..589B}
R.~{Belmont}.
\newblock {Numerical computation of isotropic Compton scattering}.
\newblock {\em Astronomy and Astrophysics}, 506:589--599, November 2009.

\bibitem{1976oup..book.....H}
G.~{Hall} and J.~M. {Watt}.
\newblock {\em {Modern Numerical Methods for Ordinary Differential Equations}}.
\newblock New York, Oxford University Press, 1976.

\end{thebibliography}

\end{document}